\documentclass[%
reprint,
nobibnotes,
amsmath,amssymb,
aps,
prb,
]{revtex4-2}

\usepackage{graphicx}
\usepackage{dcolumn}
\usepackage{bm}
\usepackage{float}
\usepackage{xcolor}
\usepackage[normalem]{ulem}

\usepackage{hyperref}
\hypersetup{colorlinks = true, linkcolor=magenta,citecolor=blue, urlcolor=magenta, bookmarksnumbered = true}

\begin{document}

\title{Two-lifetime model for the cuprates revisited}

\author{Franti\v{s}ek Herman, Lucia Gelenekyov\'{a}, Hana Havranov\'{a}, and Richard Hlubina}

\affiliation{Department of Experimental Physics, Comenius University,
  Mlynsk\'{a} Dolina F2, 842 48 Bratislava, Slovakia}

\date{\today}

\begin{abstract}
Several models of the strange-metal state of the cuprate superconductors postulate the existence of strong inelastic forward scattering of the electrons, but direct evidence of such scattering is missing. Here we show that angle-resolved photoemission spectroscopy (ARPES) provides a unique tool which can address this issue. We propose a two-lifetime phenomenological model of the superconducting state of the cuprates and we show that it explains several salient low-energy features of the measured ARPES spectra. The model  enables discrimination between forward- and large-angle scattering and, in addition, gives access to the magnitude of the gap function away from the Fermi surface.
\end{abstract}
\maketitle

\section{Introduction}
After nearly four decades of intensive studies, the cause of the anomalies observed in the non-superconducting state of the cuprate superconductors remains mysterious \cite{Keimer15,Varma20}. Strong forward scattering is often considered to be responsible for the observed anomalous behavior \cite{Anderson96,Lee92,Abrikosov94,Kulic94,Grilli94,Cappelluti96,Varma06,Wang19,Phillips20,Kukreja26}. But, by definition, the effect of such scattering on transport is strongly reduced. Does it mean that forward scattering is not directly observable? In this paper we show that this is not the case, since angle-resolved photoemission spectroscopy (ARPES) of the superconducting state provides a unique tool for distinguishing between forward and large-angle scattering in the cuprates.

To illustrate the general idea in a controlled way, we consider the case of low temperatures $T$, when impurity scattering should dominate. In an anisotropic singlet superconductor, generic impurity scattering is pair-breaking \cite{Millis88}. However, the special case of forward- and $2k_F$-scattering is pair-conserving \cite{Millis88} and obeys Anderson's theorem \cite{Anderson59}. One can argue that a large part of impurity scattering in the cuprates is of forward-scattering type \cite{Abrahams00}. The reason is simple: the CuO$_2$ planes are held together by strong covalent bonds, and therefore they are well protected against defects. On the other hand, outside these planes, disorder is generically present.  Simple estimates lead then to the conclusion that out-of-plane disorder has to generate forward scattering \cite{Abrahams00}.

Thus, in order to phenomenologically describe the low-$T$ superconducting state of the cuprates, the BCS theory has to be generalized at the very least by including both types of impurity scattering, pair-conserving (due to forward scattering) and pair-breaking (due to large-angle scattering). In case of isotropic $s$-wave superconductors, a convenient two-lifetime generalization of the BCS theory has been worked out recently \cite{Herman16}. The resulting electronic Green's function, leading to the Dynes formula \cite{Dynes78} for the density of states, is given by an explicit expression with several favourable properties \cite{Herman17a}: it is analytic in the upper half-plane, it has correct large-energy asymptotics, its spectral function is positive definite, and it does not break the particle-hole symmetry. Within this theory a superconductor is characterized by three energy scales: superconducting gap $\Delta$, pair-breaking scattering rate $\Gamma$, and pair-conserving scattering rate $\Gamma_s$. Only the parameters $\Delta$ and $\Gamma$ enter the Dynes formula \cite{Dynes78}.

In order to describe anisotropic superconductors such as the cuprates, here we make use of what we will call momentum-resolved Dynes phenomenology (MRDP).  Within MRDP, we postulate that for every momentum ${\bf k}$, the energy dependence of the electron Green's function is described by the theory of Dynes superconductors \cite{Herman16,Herman17a} with ``local'' values of the parameters $\Delta({\bf k})$, $\Gamma_s({\bf k})$, and $\Gamma({\bf k})$. We will demonstrate that MRDP is a suitable framework for description of the available ARPES data for the cuprates, thereby giving us a unique tool for a direct measurement of the magnitude of forward and large-angle scatterings.

The idea that a two-lifetime phenomenology is necessary in order to describe the superconducting state of the cuprates is not new. In fact, in \cite{Reber12} the authors introduced the concept of the tomographic density of states (TDoS) and showed that TDoS can be described by the Dynes formula. The Dynes parameter $\Gamma$ turned out to be much smaller than the quasiparticle width. Moreover, the magnitude of $\Gamma$ so defined turned out to be comparable with the energy smearing observed in tunneling spectroscopy \cite{DeCastro08}. Building on this observation in \cite{Kondo15} the authors argued, by means of an analysis of spectral functions at $k=k_F$, that two lifetimes are necessary to fit the data. However, the analysis in \cite{Kondo15} was based on an unphysical Norman's formula (NF) for the Green's function proposed in \cite{Norman98}. Moreover, the interpretation of the two lifetimes was unclear.

Here we present a substantially improved continuation of the program initiated in \cite{Reber12,Kondo15}. In Section~\ref{sec:MRDP}, we start by introducing the basic formulae of MRDP and show that MRDP enables a clear interpretation of the physical meaning of the two lifetimes. In Section~\ref{sec:MRDP_Cuprates}, we demonstrate that MRDP explains the experimentally observed low-energy features of the measured ARPES spectra in the cuprates. We find that the necessity of introducing two lifetimes is much more evident in different aspects of ARPES than in the $k=k_F$ data. In Section~\ref{sec:Comparison}, we compare  MRDP with previous work. In particular, we demonstrate that within MRDP the shortcomings of the widely used NF formalism \cite{Norman98} do not appear. Finally, in Section~\ref{sec:Conclusion} we conclude. Several more technical aspects of our work are relegated to the Appendices.

\section{Momentum-resolved Dynes phenomenology (MRDP)}\label{sec:MRDP}
The goal of this Section is to introduce a formula for the Green's function $G({\bf k},\omega)$ of an electron with momentum ${\bf k}$ and energy $\omega$ in the superconducting state of the cuprates. Let us start by noting that, for any fixed momentum ${\bf k}$, the Green's function $G({\bf k},\omega)$ as a function of $\omega$ has to satisfy several constraints: (i) it has to be analytic, (ii) it has to exhibit correct large-energy asymptotics, and (iii) its spectral function should be positive definite. Moreover, we require that the Green's function should: (iv) exhibit particle-hole symmetry, (v) discriminate between pair-conserving and pair-breaking scattering, (vi) reduce to the BCS formula in absence of scattering, (vii) lead to a density of states which depends only on the pair-breaking rate, (viii) not discriminate between the two types of scattering in the normal state. The simplest theory which satisfies all these requirements is the Dynes phenomenology introduced in Refs.~\cite{Herman16,Herman17a}. MRDP is just the ${\bf k}$-resolved version of that phenomenology. For a heuristic derivation of MRDP for a $d$-wave superconductor, see Appendix~\ref{appendix_heuristics}.

The MRDP Green's function of an electron with momentum ${\bf k}$ and energy $\omega$, derived in Refs.~\cite{Herman16,Herman17a}, can be rewritten in a BCS-like form:
\begin{equation}
 G({\bf k},\omega)=
\frac{U^2(\omega)}{\Omega+i\Gamma_s-\varepsilon_{\bf k}}
+\frac{V^2(\omega)}{\Omega+i\Gamma_s+\varepsilon_{\bf k}}.
\label{eq:green_f}
\end{equation}
For the sake of brevity, if no confusion arises, we do not indicate the ${\bf k}$-dependence of $\Delta$, $\Gamma_s$, and $\Gamma$. In Eq.~\eqref{eq:green_f} we have introduced a complex $\omega$-dependent energy scale $\Omega(\omega)=\sqrt{(\omega+i\Gamma)^2-\Delta^2}=\Omega_1+i\Omega_2$, where the square root is taken so that  $\Omega_1(\omega)$ is an odd function of $\omega$, while $\Omega_2(\omega)$ is even and positive. Explicitly, we take
\begin{eqnarray*}
\Omega_1&=&{\rm sgn}(\omega)\{[(z^2+4\omega^2\Gamma^2)^{1/2}+z]/2\}^{1/2}, 
\\
\Omega_2&=&\{[(z^2+4\omega^2\Gamma^2)^{1/2}-z]/2\}^{1/2},
\end{eqnarray*} 
with $z=\omega^2-\omega_\ast^2$ and $\omega_\ast=\sqrt{\Delta^2+\Gamma^2}$. We have also introduced the following $\omega$-dependent weight factors:
$$
U^2(\omega)=\frac{\omega+i\Gamma+\Omega}{2\Omega},
\qquad
V^2(\omega)=\frac{\omega+i\Gamma-\Omega}{2\Omega}.
$$
Note that $U^2(\omega)$, $V^2(\omega)$, and $\Omega(\omega)$ depend only on the pair-breaking rate $\Gamma$.

The electron spectral function is given by the usual expression $A({\bf k},\omega)=-{\rm Im}\left[G({\bf k},\omega)\right]/\pi$.  When the pair-conserving scattering rate $\Gamma_s$ vanishes, it simplifies to
\begin{equation}
A({\bf k},\omega)=
u_{\bf k}^2\delta_\Gamma(\omega-E_{\bf k})+v_{\bf k}^2\delta_\Gamma(\omega+E_{\bf k}),
\label{eq:spectral_no_gammas}
\end{equation}
where the symbol $\delta_\Gamma(x)$ stands for a Lorentzian with width $\Gamma$, $\delta_\Gamma(x)\equiv \pi^{-1}\Gamma/(x^2+\Gamma^2)$. This means that the spectral function consists of just two peaks at energies $\pm E_{\bf k}$ where $E_{\bf k}=\sqrt{\varepsilon_{\bf k}^2+\Delta^2}$ is the BCS quasiparticle energy, with the BCS weight factors $u_{\bf  k}^2=(E_{\bf k}+\varepsilon_{\bf k})/(2E_{\bf k})$ and $v_{\bf  k}^2=(E_{\bf k}-\varepsilon_{\bf k})/(2E_{\bf k})$. In the limit $\Gamma\rightarrow 0$, Eq.~\eqref{eq:spectral_no_gammas} further simplifies to the standard BCS result.

Conversely, if the pair-conserving scattering rate $\Gamma_s$ is finite but $\Gamma=0$, the spectral function is strictly zero inside the gap, i.e. for $|\omega|<\Delta$, while for $|\omega|>\Delta$ we find
\begin{equation}
A({\bf k},\omega)=
U^2(\omega)\delta_{\Gamma_s}(\Omega_1-\varepsilon_{\bf k})
+V^2(\omega)\delta_{\Gamma_s}(\Omega_1+\varepsilon_{\bf k}).
\label{eq:spectral_no_gamma}
\end{equation}
Since in this case $\Omega_1=\rm{sgn}(\omega)\sqrt{\omega^2-\Delta^2}$, the Lorentzians again  exhibit peaks at $\omega=\pm E_{\bf k}$. The spectral function contains also two other peaks at $\omega=\pm\Delta$, where the prefactors $U^2(\omega)$ and $V^2(\omega)$ exhibit square-root singularities. 

In the general case when both $\Gamma$ and $\Gamma_s$ are finite,  for $\varepsilon_{\bf k}\neq 0$ the spectral function again exhibits a four-peak structure with the peaks located at energies $\pm E_{\bf  k}$ and $\pm\Delta$, but the singularities at $\pm\Delta$ predicted by Eq.~\eqref{eq:spectral_no_gamma} are smeared by the finite value of $\Gamma$, see Ref.~\cite{Herman17a} for details.

We stress that one can discriminate between $\Gamma_s$ and $\Gamma$ only in the superconducting state. When $\Delta=0$, $A({\bf k},\omega)$ is a Lorentzian with width $\Gamma_{\rm tot}=\Gamma_s+\Gamma$, thus only the total scattering rate $\Gamma_{\rm tot}$ can be extracted from $A({\bf k},\omega)$.

\begin{figure}[t]
\includegraphics[width = 5cm]{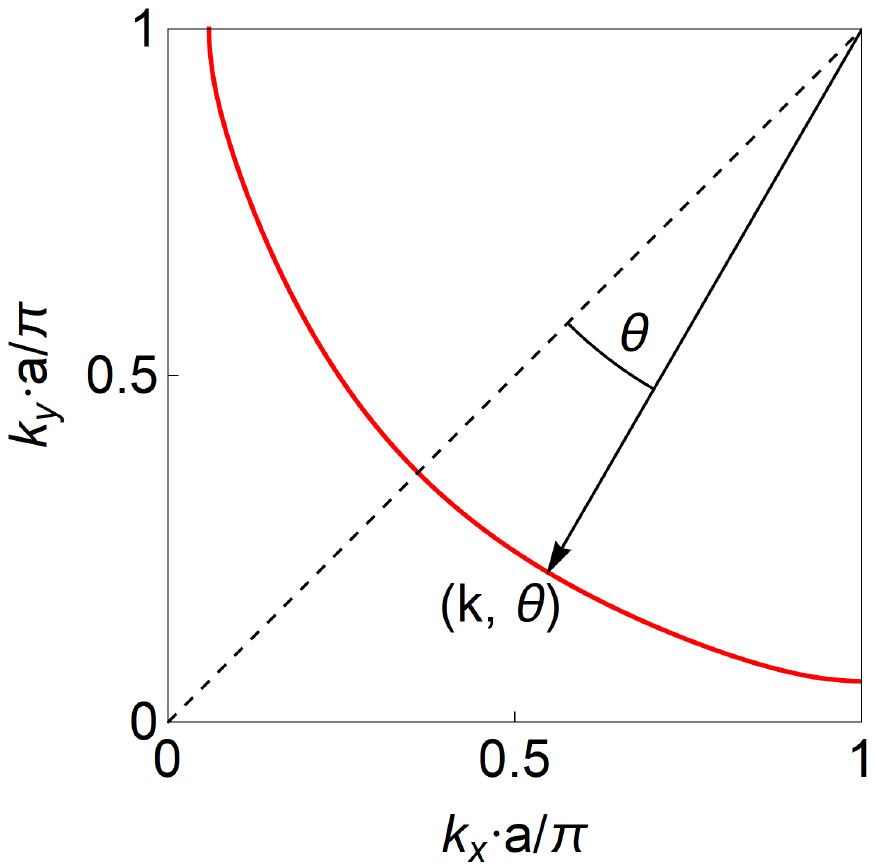}
\caption{Representation of momentum ${\bf k}$ inside the Brillouin zone in terms of the angle $\theta$ and radial coordinate $k$. The red line shows the Fermi surface of the studied model.}
\label{fig:schematic}
\end{figure}

In this paper MRDP will be applied to several aspects of the measured low-energy ARPES spectra in the cuprates \cite{Reber12,Kondo15,Chen22,Reber15,Ye24,Shah25,Bok16,Li18,Yan23}. We consider the simplest normal-state electron dispersion capturing the gross features of the cuprates, $\varepsilon_{\bf k} = -2 t \big(\cos k_xa +\cos k_ya \big) + 4t'\cos k_xa \cos k_ya - \mu$. Here $a$ is the lattice constant, $t =0.335$~eV and $t' = 0.125$~eV are hopping amplitudes, and $\mu = -0.48$~eV is the chemical potential. The superconducting state is described by the $d$-wave gap 
\begin{equation}
\Delta_{\bf k} = \Delta_d \big(\cos k_xa -\cos k_ya \big)/2.
\label{eq:gap_function}
\end{equation}
In what follows, we often parametrize the two-dimensional momentum ${\bf k}$ by $(k,\theta)$, where the angle $\theta$ defines the tomographic cut, see Fig.~\ref{fig:schematic}, and $k$ measures the position along the cut.

\begin{figure}[t]
\includegraphics[width = 8.7cm]{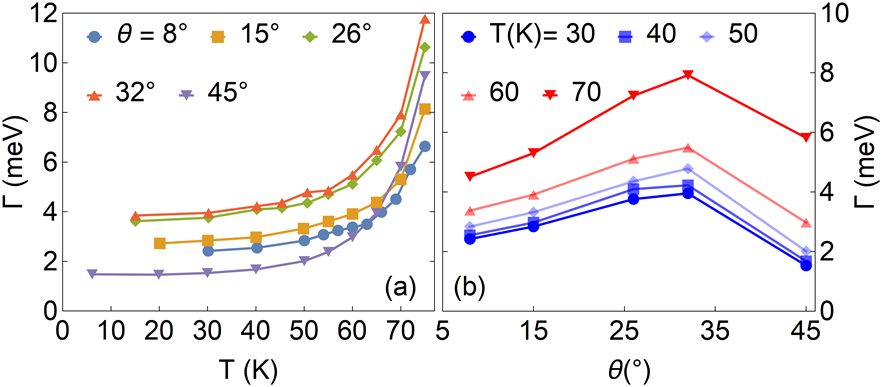}
\caption{Pair-breaking scattering rate $\Gamma$, as determined from the tomographic density of states reported in~\cite{Chen22}.}
\label{fig:tdos}
\end{figure}

\section{Application of MRDP to the cuprates}\label{sec:MRDP_Cuprates}
\subsection{Tomographic density of states}
In the rest of this paper we apply MRDP to several salient low-energy features of the measured ARPES spectra in the cuprates. We start with the tomographic density of states, TDoS, which can be found by integrating the spectral functions along a fixed "tomographic" cut perpendicular to the Fermi surface, $N(\omega,\theta)=\int d k A(k,\theta,\omega)$  \cite{Reber12}.  Starting with Eq.~\eqref{eq:green_f} and neglecting the (presumably) weak dependence of $\Delta(k,\theta)$, $\Gamma(k,\theta)$, and $\Gamma_s(k,\theta)$ on the radial coordinate $k$, we find that $N(\omega,\theta)$ is within MRDP proportional to the Dynes formula 
\begin{equation}
n(\omega,\theta)={\rm Re}\left[\frac{\omega+i\Gamma}{\sqrt{(\omega+i\Gamma)^2-\Delta^2}}\right],
\label{eq:Dynes_formula}
\end{equation} 
in agreement with the experimental results \cite{Reber12}. Note that TDoS does not depend on the value of $\Gamma_s(\theta)$, but only on the local values of $\Gamma(\theta)$ and $\Delta(\theta)$. It is this feature of MRDP which explains the large difference between the lifetimes observed in ARPES and in STM.

In Ref.~\cite{Chen22}, the authors report the temperature- and angle-dependence of the superconducting gap $\Delta(T,\theta)$, as well as the angle-dependent ratio $\alpha(T,\theta)=A_0(T,\theta)/A_0(140~K,\theta)$, where $A_0(T,\theta)$ is the momentum-integrated zero-energy intensity at temperature $T$ and angle $\theta$. Here we assume that $\alpha$ measures the normalized TDoS, and therefore within MRDP we estimate that $\alpha=\Gamma/\sqrt{\Delta^2+\Gamma^2}$. From the reported values of $\Delta(T,\theta)$ and $\alpha(T,\theta)$ we determine $\Gamma(T,\theta)$ and the results are plotted in Fig.~\ref{fig:tdos}. We find that $\Gamma$ is roughly isotropic and, upon approaching $T_c$, it becomes strongly $T$-dependent.

These results for $\Gamma(T,\theta)$ are in reasonable agreement with the roughly isotropic pair-breaking scattering rate $\Gamma\sim \Delta_d/10\approx 3$~meV observed previously in the near-nodal regime of optimally doped Bi2212 in Ref.~\cite{Reber12}. Overdoped samples of Bi2212 also exhibit $\Gamma\sim \Delta_d/10\approx 3$~meV in the near-nodal region at low temperatures \cite{Reber15}. 

We would like to point out that the values of $\Gamma(T,\theta)$ extracted from TDoS (and thus from ARPES) are fully compatible with the finding that the low-$T$ tunneling data at $|\omega|\lesssim \Delta_d$ can be fitted by a simple $d$-wave form with a comparably small smearing $\Gamma$, see e.g. \cite{DeCastro08}. This provides a strong indication that the ARPES data are not contaminated by an energy resolution lower than that of the tunneling.

\subsection{Spectral functions at the Fermi surface}
\begin{figure}[b]
\includegraphics[width = 8.7 cm]{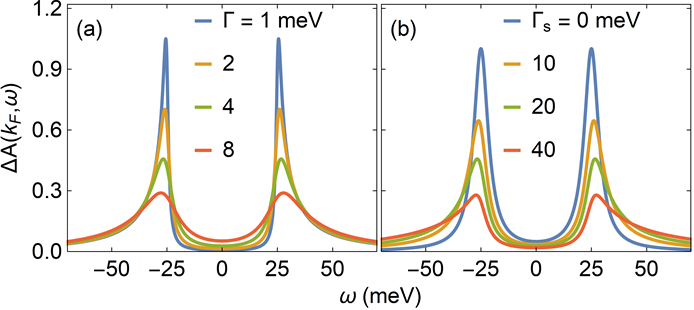}
\caption{Dependence of the Fermi-surface spectral function $A(k_F,\omega)$ for $\Delta=25$~meV on the scattering rates. (a): Fixed $\Gamma_s=20$~meV. Increasing $\Gamma$ symmetrically broadens the peaks. (b): Fixed $\Gamma=4$~meV. Increasing $\Gamma_s$ generates asymmetric shape of the peaks, shifting weight from $|\omega|<\omega_\ast$ to $|\omega|>\omega_\ast$.}
\label{fig:schem}
\end{figure}

The MRDP parameters $\Delta$, $\Gamma_s$, and $\Gamma$ can be determined most straightforwardly by fitting the ARPES spectra at the Fermi surface $k=k_F$, since in that case the spectra do not depend on the energy dispersion $\varepsilon_{\bf k}$.  At the Fermi surface, the spectral function is given by
\begin{equation}
A(k_F,\omega)=\frac{1}{\pi}\frac{\Gamma(\omega^2+\omega_\ast^2)+\Gamma_s(\omega \Omega_1+\Gamma\Omega_2)}{(\Omega_1^2+\Omega_2^2)\left[\Omega_1^2+(\Omega_2+\Gamma_s)^2\right]},   
\end{equation}
where $\omega_\ast=\sqrt{\Delta^2+\Gamma^2}$.  The function $A(k_F,\omega)$ is even and it exhibits two peaks at $\omega\approx\pm\omega_\ast$. Figure~\ref{fig:schem} shows that the two scattering rates modify the spectral function in different ways: The pair-breaking rate $\Gamma$ symmetrically broadens the peaks, while the forward-scattering rate $\Gamma_s$ generates a finite asymmetry of the peaks, increasing their weight predominantly for $|\omega|>\omega_\ast$. Note that the shape of $A(k_F,\omega)$ is very sensitive even to small values of $\Gamma$.

The MRDP parameters were obtained by fitting the data from Ref.~\cite{Kondo15} for optimally doped Bi2212. The background is taken as negligible in the fits of the ARPES data. Thus, the fits depend on four parameters: $\Delta$, $\Gamma_s$, $\Gamma$, and the normalization factor. The fits were performed for energies in the range $|\omega|<\Lambda$, where $\Lambda=35$~meV. Larger values of $|\omega|$ were not considered in the fits, because scattering can not any more be taken as elastic outside this energy range. This value of the cutoff $\Lambda$  agrees with the estimate which makes use of the tunneling data \cite{DeCastro08}. In fact, the low-temperature density of states of the cuprates can be fitted by the BCS formula for a $d$-wave superconductor with a small and constant (energy- and angle-independent) Dynes parameter $\Gamma$. Such fits are known to be very good at energies $|\omega|$ ranging up to slightly above the coherence peaks of the density of states, see e.g. Ref.~\cite{DeCastro08}.

\begin{figure}[t!]
\includegraphics[width = 8.6 cm]{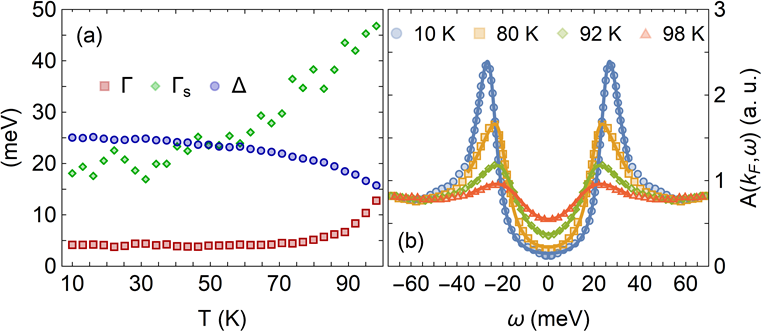}
\caption{(a): Temperature dependence of the MRDP parameters obtained by fitting the spectral functions at $k=k_F$ and $\theta=24^\circ$ in optimally doped Bi2212, taken from Ref.~\cite{Kondo15}. In agreement with Refs.~\cite{Reber12,Kondo15,Chen22,Li18}, the gap and the scattering rates are continuous across the critical temperature $T_c=92$~K. (b): Spectral functions  (symbols) and their MRDP fits (lines) for selected temperatures.}
\label{fig:quantitative}
\end{figure}

As an example, in Fig.~\ref{fig:quantitative} we plot the MRDP parameters obtained from Ref.~\cite{Kondo15} at angle $\theta=24^\circ$.  At low temperatures up to $T\approx 50$~K both extracted scattering rates are roughly $T$-independent, as expected for elastic processes. We find that $\Gamma_s\gg \Gamma$, in  agreement with the picture proposed in Ref.~\cite{Abrahams00}. Strictly speaking, since MRDP is motivated by elastic scattering, our results are well controlled only in the low-energy, low-temperature sector. However, it has been found recently that, at least in the context of the so-called power-law liquids \cite{Reber19}, for anomalous scattering mechanisms the $T$-dependence of the electron lifetime is much stronger than the $\omega$-dependence \cite{Skrlec24}. This suggests that the $T$-dependence of the scattering parameters $\Gamma$ and $\Gamma_s$ might be determined correctly, even if the $\omega$-dependence of the self-energy is borrowed from a theory with only elastic scattering. Strong support for such modeling comes from the fact that, even for $T\gtrsim 50$~K, MRDP does fit the data at $|\omega|<\Lambda$ quite well. Note that the same argument may justify many previous analyses which consider only the $T$-dependence and neglect the $\omega$-dependence of the parameters entering the theory, in particular also the NF-based approach used in Ref.~\cite{Kondo15}. Obviously, further work is needed to put this argument on a more solid ground.

In Figs.~\ref{fig:leporelo1},~\ref{fig:leporelo2} in the Appendix~\ref{appendix_fermi_surface} we show the results of the MRDP analysis for all angles studied in Ref.~\cite{Kondo15}. As one can see there, the overall quality of the fits is good and it improves with increasing $\theta$. The $T$-dependence of the extracted parameters is  for all angles similar to the results for $\theta=24^\circ$. 

\begin{figure*}[t!]
\includegraphics[width = 14.9cm]{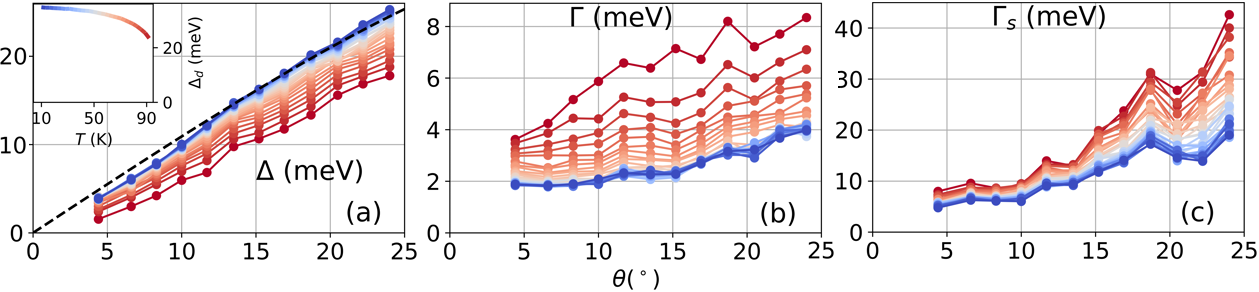}
\caption{Angular dependence of the MRDP parameters extracted in Fig.~\ref{fig:leporelo2} for temperatures between 10~K (dark blue) and 92~K (dark red). For all temperatures, the gap in (a) can be fitted by the $d$-wave formula Eq.~\eqref{eq:gap_function}. The dashed line corresponds to $T=10$~K and $\Delta_d=34.8$~meV. The inset in (a) shows the temperature dependence of $\Delta_d$.}
\label{fig:uhly}
\end{figure*}

The angular dependence of the extracted parameters is shown in Fig.~\ref{fig:uhly}. For all studied temperatures, the gap $\Delta_{\bf k}$ can be fitted by the $d$-wave formula Eq.~\eqref{eq:gap_function}. The extracted parameter $\Delta_d(T)$ is a decreasing function of temperature, as expected, but it is finite even above the critical temperature $T_c=92$~K, in agreement with the findings of Ref.~\cite{Kondo15}. 

Note that both the angular- and the temperature-dependence of the pair-breaking scattering rate $\Gamma$ agree quite well with the results obtained from the TDoS data in Ref.~\cite{Chen22}, see Fig.~\ref{fig:tdos}, and also from Refs.~\cite{Reber12,Reber15}. We identify $\Gamma$ with the isotropic scattering rate $\Gamma_{\rm iso}$ found by magnetotransport \cite{Grissonnanche21}. In fact, close to $T_c$, our magnitude of $\Gamma\approx 4-8$~meV (see Figs.~\ref{fig:tdos}, \ref{fig:uhly}) roughly agrees with $\Gamma_{\rm iso}\approx 0.6 k_BT$ found in \cite{Grissonnanche21}. 

The forward scattering rate $\Gamma_s$ is much larger than $\Gamma$ for all studied angles, in agreement with expectations. We believe that $\Gamma_s$ and the anisotropic scattering rate  $\Gamma_{\rm aniso}$ found in Refs.~\cite{Grissonnanche21,Kaminski05} are caused by the same mechanism, since both exhibit a similar dependence on the angle $\theta$. At $T_c$ we find for our largest angle $\theta=24^\circ$ that $\Gamma_s\approx 42$~meV, see Fig.~\ref{fig:uhly}. This is larger than $\Gamma_{\rm aniso}\approx 20$~meV found by magnetotransport \cite{Grissonnanche21} in the maximum at $\theta=45^\circ$. The finding that $\Gamma_s(\theta)\gg\Gamma_{\rm aniso}(\theta)$ from Ref.~\cite{Grissonnanche21} is of course consistent with the forward-scattering origin of $\Gamma_s$, as realized already in Ref.~\cite{Abrahams00}. 

Surprisingly, we find that $\Gamma_s$ is $T$-dependent at $T\lesssim T_c$: this is different from the normal-state result that $\Gamma_{\rm aniso}$ is independent of temperature \cite{Grissonnanche21} and energy \cite{Kaminski05}. We hypothesize that this $T$-dependence persists also at temperatures above $T_c$. The reason is the following: According to normal-state ARPES \cite{Kaminski05}, the total scattering rate at the Fermi surface in optimally doped Bi2212 at 140~K, for instance at angle $\theta=24^\circ$, is $\Gamma_{\rm tot}\approx 160$~meV. This is larger than $\Gamma_{\rm tot}=\Gamma_s+\Gamma\approx 50$~meV extracted using MRDP at $T_c$ from the data in \cite{Kondo15} at the same angle. The pair-breaking rate at $T_c$ and $\theta=24^\circ$ is only $\Gamma\approx 8$~meV, suggesting that the growth of $\Gamma_{\rm tot}$ between $T_c$ and 140~K must be largely due to an increase of $\Gamma_s$. Therefore the forward-scattering rate $\Gamma_s$ above $T_c$ has to be caused, to a substantial part, by inelastic scattering affecting mostly anti-nodal electrons.


\subsection{Momentum maps}
\begin{figure}[h!]
\includegraphics[width = 8.6 cm]{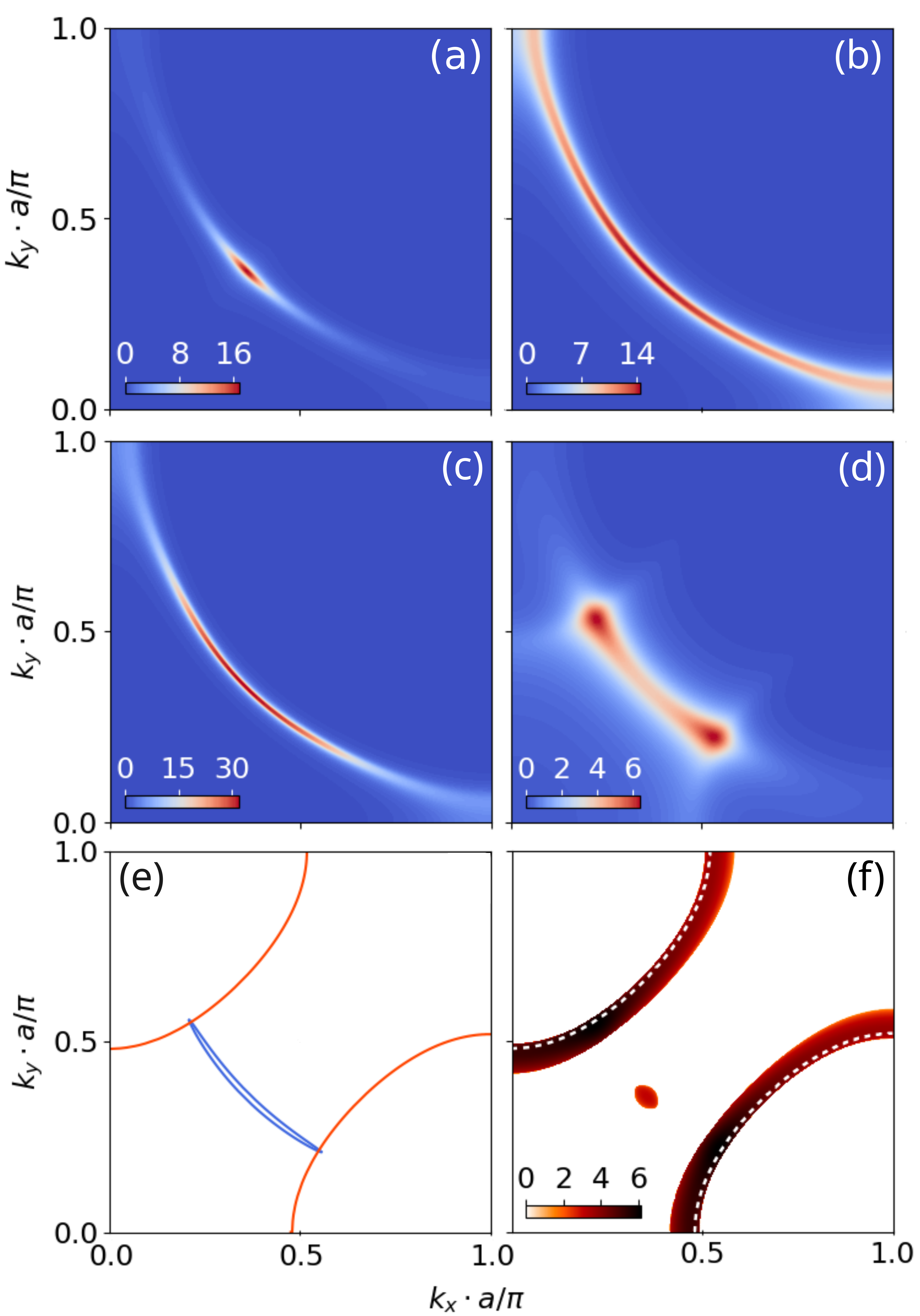}
\caption{(a-d): Momentum maps of $A({\bf k},\omega)$  predicted by MRDP for model superconductors with momentum-independent scattering rates $\Gamma$ and $\Gamma_s$. (a,b) shows data at the chemical potential $\omega=0$ for $\Gamma_s=16$~meV. (a): $\Delta_d=30$~meV and $\Gamma=3$~meV.  (b): $\Delta_d=6$~meV and $\Gamma=6$~meV. (c,d) shows data at finite energy $\omega=-14$~meV for $\Delta_d=30$~meV. (c): $\Gamma=10$~meV and $\Gamma_s=0$. (d): $\Gamma=2$~meV and $\Gamma_s=80$~meV. (e): The banana-shaped line where $E_{\bf k}=-\omega$ (blue line) and the gap arcs where $|\Delta_{\bf k}|=-\omega$ (red lines) for the same parameters as in (c,d). (f): Momentum map of the logarithm of the second derivative of the data in (d) with respect to energy, $\log[-A''({\bf k},\omega)]$. Only points where $A''({\bf k},\omega)<0$ are shown. White dashed lines show the gap arcs.}
\label{fig:momentum_maps}
\end{figure}

Next we study the spectral function $A({\bf k},\omega)$ for fixed $\omega$ as a function of momentum ${\bf k}$ in the 2D Brillouin zone, to be called the momentum map. In overdoped but still superconducting Bi2201 it was recently found that the momentum map at the chemical potential, i.e. for $\omega=0$, resembles that of a normal metal \cite{Ye24}. This result may be explained by the single-lifetime theory leading to Eq.~\eqref{eq:spectral_no_gammas} as in Refs.~\cite{Norman07,Chubukov07}, but it is consistent also with the full MRDP. In fact, for $\omega=0$ Eq.~(\ref{eq:green_f}) implies that 
\begin{equation}
A({\bf k},0) = w(\theta) \delta_{\Gamma_{\rm eff}}(\varepsilon_{\bf k}).
\label{eq:mdf}
\end{equation}
Thus, the $\omega=0$ momentum map consists of Lorentzians  along tomographic cuts with maxima at the normal-state Fermi surface $\varepsilon_{\bf k}=0$, with angle-dependent widths $\Gamma_{\rm eff}(\theta) = \Gamma_s+\omega_\ast$ and weights $w(\theta)=\Gamma/\omega_\ast$.  Momentum maps calculated using Eq.~\eqref{eq:mdf} in the Brillouin zone of a model cuprate superconductor are shown in Fig.~\ref{fig:momentum_maps}. For $\Gamma\ll \Delta_d$ (Fig.~\ref{fig:momentum_maps}a), the spectral function is finite only in the vicinity of the nodal point, as expected. However, for a pair-breaking rate $\Gamma$ comparable to $\Delta_d$ (Fig.~\ref{fig:momentum_maps}b), the momentum map resembles the experimental results of \cite{Ye24}.

Next we consider momentum maps at energy $\omega<0$. In a BCS superconductor with vanishing scattering rates $\Gamma$ and $\Gamma_s$, the momentum map  is finite only at those ${\bf k}$-points where $E_{\bf k}=-\omega$, see Fig.~\ref{fig:momentum_maps}e. Finite values of both, $\Gamma$ and $\Gamma_s$, contribute to the broadening of these so-called bananas. However, the foward-scattering rate $\Gamma_s$ produces an additional effect: it generates a finite signal in the momentum map also along the "gap arcs" where $|\Delta_{\bf k}|=-\omega$,  also shown in Fig.~\ref{fig:momentum_maps}e. As a result, the spectral function in the ends of the bananas (i.e., at angle $\theta_0$) is enhanced with respect to their interior, as predicted theoretically long ago by Markiewicz \cite{Markiewicz04}. The need for the existence of an enhancement mechanism of some sort in the cuprates has been stressed also in later work \cite{Kim10}.

More quantitatively, the maximal spectral function (as a function of $k$) in the nodal direction is $A_{\rm norm}=1/(\pi\Gamma_{\rm tot})$. We estimate $\theta_0$ as the angle where $\omega^2=\Delta(\theta_0)^2+\Gamma(\theta_0)^2$ holds. The ratio between $A_{\rm norm}$ and the spectral function in the bright spots at the end of the bananas, $A_{\rm bright}=A(k_F,\theta_0)$, strongly depends on the value of $\Gamma_s$. For $\Gamma_s=0$ we find $A_{\rm bright}/A_{\rm norm}=1/2$, whereas for $\Gamma_s\gg\Delta\gg \Gamma$, we estimate $A_{\rm bright}/A_{\rm norm}\approx \sqrt{\Delta/(4\Gamma)}\gg 1$. 

This effect is demonstrated explicitly in Figs.~\ref{fig:momentum_maps}c,d, where we compare momentum maps for $\Gamma_s=0$ and $\Gamma_s\neq 0$. The parameter values used in Figs.~\ref{fig:momentum_maps}c,d are motivated by the very recent experimental observation of bright spots at the ends of the bananas  in moderately underdoped Bi2212 \cite{Shah25}. In particular, in order to take into account the observed quasiparticle widths, a largish value of $\Gamma=10$~meV has to be assumed, if $\Gamma_s=0$. For such a value of $\Gamma$ , bright spots at the ends of the bananas do not appear, even if the finite momentum resolution of the ARPES experiment is taken into account. For more details, see Appendix~\ref{appendix_momentum maps}.

To summarize, the data presented in Ref.~\cite{Shah25} strongly support our MRDP modeling with finite values of $\Gamma_s$. We emphasize that all maps in Fig.~\ref{fig:momentum_maps} are completely determined by the low-energy properties of the spectral functions. For an analysis of the so-called joint density of states also presented in \cite{Shah25}, see Appendix~\ref{appendix_momentum maps}.

\subsection{Tomographic maps}
Much more common in the literature than the momentum maps are 2D plots of the spectral function $A(k,\theta,\omega)$ at fixed angle $\theta=$ const, to be called tomographic maps in what follows. For recent examples of experimental tomographic cuts for the cuprates, see Refs.~\cite{Bok16,Li18,Yan23}. A commonly observed feature of tomographic cuts, which does not seem to have attracted attention so far, is that the maximal spectral weight at a fixed energy $\omega$, $A_{\rm max}(\omega)={\rm max}_k A(k,\omega)$, increases as $|\omega|$ moves towards the energy gap $|\omega|\approx \Delta$. Note that for large $|\omega|$, within MRDP $A_{\rm max}(\omega)=A_{\rm norm}$. On the other hand, for energy $\omega_\ast$ close to the gap $A_{\rm max}(-\omega_\ast)\approx A(k_F,|\omega_\ast|)=A_{\rm bright}$. Therefore, the modulation of the spectral function $A$ along the BCS line $\omega=-E_{\bf k}$ in the tomographic cut exhibits the same behavior as the modulation of $A$ along the banana in the momentum map: in both maps, one compares the magnitude of $A$ at $k=k_F$ with $A$ for momenta far away from the Fermi surface. As a result, a finite $\Gamma_s$ is required in order  that MRDP resembles the experimental tomographic maps. This is explicitly demonstrated in the numerically obtained tomographic maps presented in Fig.~\ref{fig:tomographic_cut}. For more details regarding the comparison to experiments, see Appendix~\ref{appendix_tomographic_maps}. 

It is pleasing to note that, also on the experimental side, bright spots are consistently found both in momentum maps~\cite{Shah25}, as well as in tomographic maps~\cite{Bok16,Li18,Yan23}. 

\begin{figure}[t!]
\includegraphics[width = 8.7 cm]{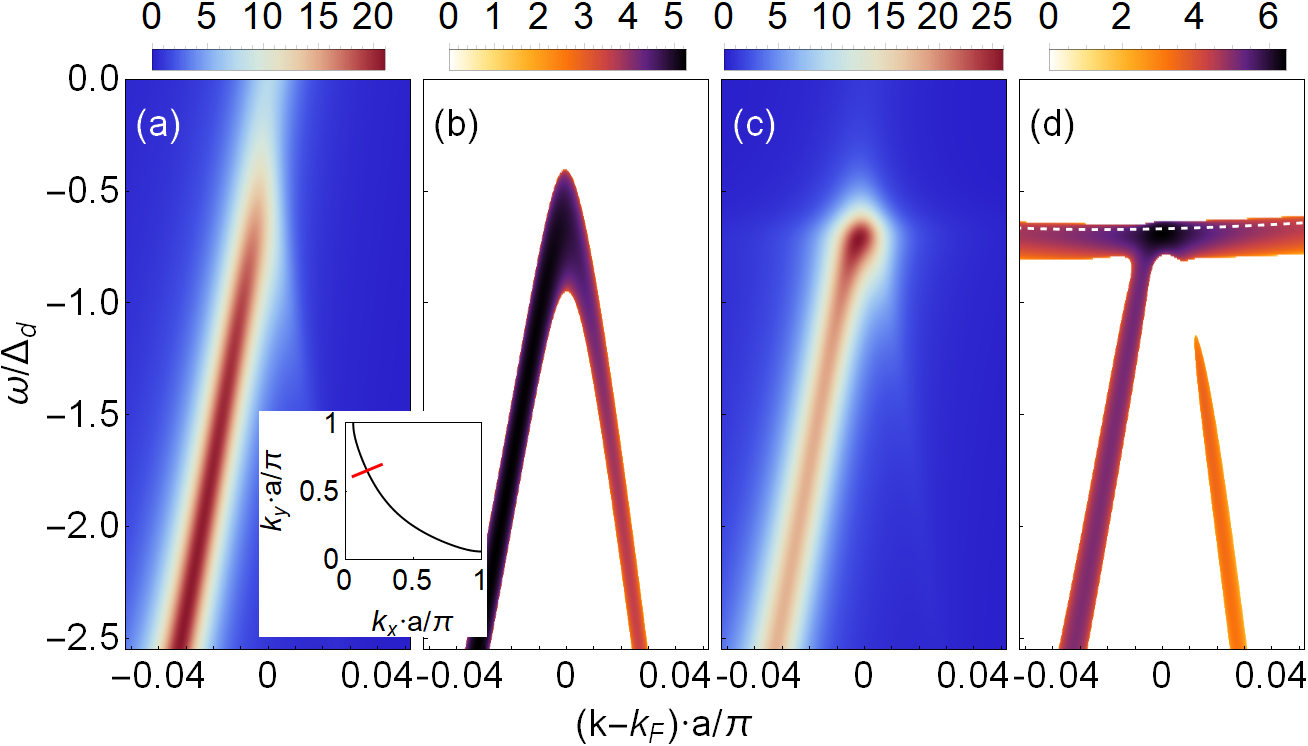}
\caption{(a,c): Tomographic maps of $A(k,\omega)$ predicted by MRDP along the path indicated in the inset for a superconductor with $\Delta_d=30$~meV. (b,d): Logarithm of the second derivatives with respect to energy $\log[-A''(k,\omega)]$ of the same data. Only points where $A''(k,\omega)<0$ are shown. (a,b): $\Gamma=15$~meV and $\Gamma_s=0$~meV. (c,d): $\Gamma=3$~meV and $\Gamma_s=15$~meV. White dashed line in (d): the function $\omega=-\Delta_{\bf k}$ along the tomographic cut.}
\label{fig:tomographic_cut}
\end{figure}

\subsection{Gap arcs}
Having presented a set of arguments strongly supporting the applicability of MRDP to the cuprates, in the rest of this paper we propose that ARPES offers, in addition,  a unique possibility to measure the so-far unexplored momentum dependence of the gap function $\Delta_{\bf k}$ also away from the Fermi surface. Such information, if available, would provide an additional constraint on the acceptable pairing mechanism in the cuprates: for instance, the functional form Eq.~\eqref{eq:gap_function} implies that Cooper pairs are formed on the nearest-neighbor links of the CuO$_2$ plane.

\begin{figure}[b!]
\includegraphics[width = 8.4 cm]{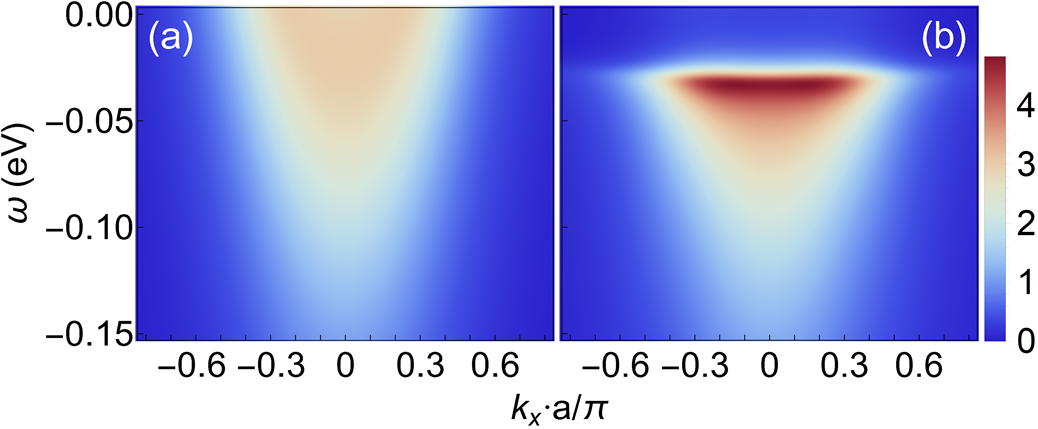}
\caption{Tomographic maps of $A(k,\omega)$ predicted by MRDP along the antinodal cut $\theta=45^\circ$ for $T$-independent scattering rates $\Gamma=5$~meV and $\Gamma_s=100$~meV. (a): Normal state. (b): Superconducting state with $\Delta_d=30$~meV.}
\label{fig:antinode}
\end{figure}

To this end, let us reconsider Fig.~\ref{fig:momentum_maps}d, where we plot the momentum map of $A({\bf k},\omega)$ for large forward scattering rate $\Gamma_s$ and small pair-breaking rate $\Gamma$. Note that the gap arcs where $|\Delta_{\bf k}|=-\omega$ start to be visible in such a map, thereby providing direct access to the gap function. Later we will elaborate on the fact  that $A({\bf k},-|\Delta_{\bf k}|)$ grows with $\Gamma_s$ and decreases with $\Gamma$.

Surprisingly, the $|\Delta_{\bf k}|=-\omega$ part of the electron spectral function might have already been observed. In fact, the authors of Ref.~\cite{Yan23} have calculated the second derivative of the measured spectral function with respect to energy, $A''({\bf k},\omega)$, in order to emphasize the local maxima of the spectral function $A({\bf k},\omega)$. Besides other features, they have found large values of  $-A''({\bf k},\omega)$ at $\omega\approx -|\Delta_{\bf k}|$. Our Fig.~\ref{fig:tomographic_cut} shows that, within MRDP, taking the second derivative does strongly emphasize the $\omega=-|\Delta_{\bf k}|$ branch of the tomographic map of the spectral function. Therefore, we believe that the signal found in~\cite{Yan23} might be interpreted as a measurement of the function $\Delta_{\bf k}$ along the tomographic cut. In Fig.~\ref{fig:momentum_maps}f we show that plotting the momentum map of the second energy derivative $A''({\bf k},\omega)$ should improve the visibility of the gap arc $|\Delta_{\bf k}|=-\omega$ also in momentum maps. Alternatively, one might study maps of the ratio between superconducting and normal spectral functions.

\begin{figure*}[t]
\includegraphics[width =  0.85\textwidth]{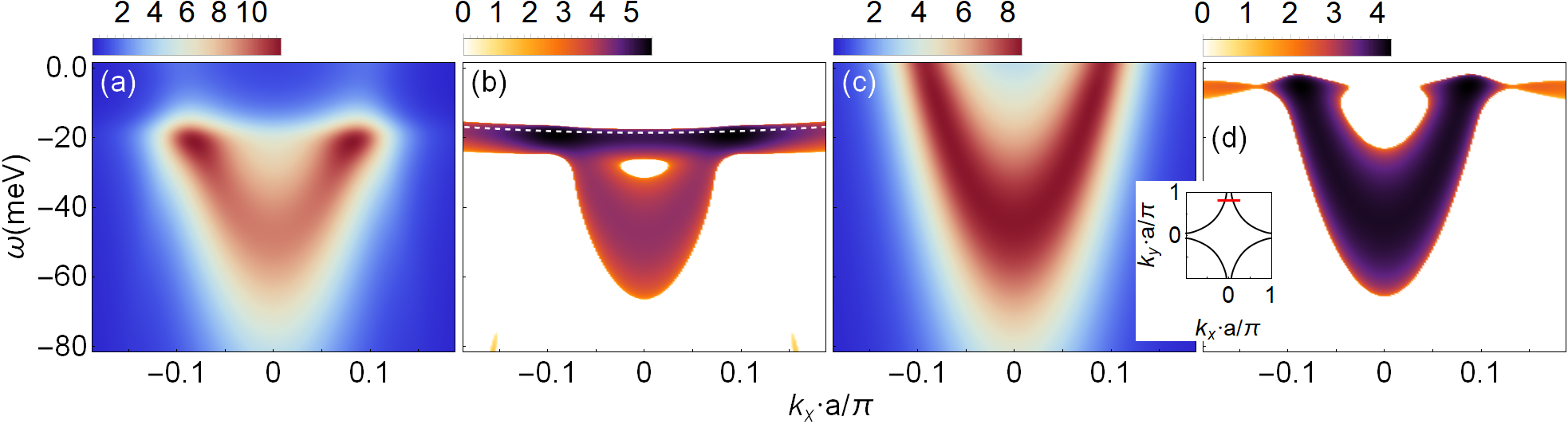}
\caption{(a,c) Spectral functions $A({\bf k},\omega)$ in the plane $(k_x,\omega)$ along the cut $k_y a= 2.6$ (shown in the inset) for a superconductor with $\Gamma_s = 30$~meV. (b,d) Logarithm of the second derivatives with respect to energy $\log[-A''(k,\omega)]$ of the data in (a,c), respectively. Only points where $A''({\bf k},\omega)<0$ are shown. White dashed line in (b): the function $\omega=-\Delta_{\bf k}$ along the studied cut. (a): $\Delta_d = 20$~meV, $\Gamma = 5$~meV relevant for Fig.~4 in \cite{Yan23} at low $T$. Bright spots at the non-interacting Fermi-surface points are clearly visible. (c): $\Delta_d = 1.5$~meV, $\Gamma = 7$~meV relevant close to $T_c$. Note that the gap arc is visible for $\Gamma_s> \Delta\gtrsim\Gamma$.}
\label{fig:antinodal_yan}
\end{figure*}

In order to explain the surprising effectivity of the second derivative method, let us consider the limit of very large forward scattering $\Gamma_s\gg \Gamma,\Delta$. In this limit, the spectral function $A({\bf k},\omega)$ for $|\omega|\ll \Gamma_s$ and arbitrary $\varepsilon_{\bf k}$ simplifies to
\begin{equation}
A({\bf k},\omega)=\delta_{\Gamma_s}(\varepsilon_{\bf k})\times
{\rm Re}\left[\frac{\omega+i\Gamma_{\bf k}}{\sqrt{(\omega+i\Gamma_{\bf k})^2-\Delta_{\bf k}^2}}\right],
\label{eq:large_gamma_s}
\end{equation}
 i.e., as a function of $\omega$, the spectral function is proportional to the Dynes formula. Since for $\Gamma_{\bf k}\ll \Delta_{\bf k}$ the latter exhibits a sharp maximum at $|\omega|\approx |\Delta_{\bf k}|$, the second derivative $A''({\bf k},\omega)$ with respect to $\omega$ is strongly negative in the vicinity of $|\omega|\approx |\Delta_{\bf k}|$ in this case, as indeed observed in Fig.~\ref{fig:momentum_maps}f. Note also that according to Eq.~\eqref{eq:large_gamma_s}, surprisingly, the superconducting gap is observable even if $\Gamma_s\gg\Delta$, provided $\Gamma\lesssim\Delta$. In the special case when $\Gamma=0$, this has been noted also previously \cite{Zhu04}. Such sharpening of the spectral function in the superconducting state is relevant in the antinodal region, and explicitly demonstrated in Fig.~\ref{fig:antinode}. 

Returning to the experimental data presented in Ref.~\cite{Yan23}, there the authors present, inter alia, high-resolution ARPES data for slightly overdoped Bi2212. They study the second derivative with respect to energy of the spectral function $A_{\rm obs}({\bf k},\omega)$ measured along several cuts in momentum space and concentrate on small features of $A_{\rm obs}''({\bf k},\omega)$ at binding energies $\approx 65$~meV and $\approx 100$~meV. However, the largest non-BCS feature of the data is present at smaller binding energies comparable to the local gap values. In particular, close to the antinode this feature appears at $\approx 20$~meV. Note that this feature can not be caused by the Fermi function $f(\omega)$ which enters the relation $A_{\rm obs}({\bf k},\omega)=f(\omega)A({\bf k},\omega)$ between the observed and theoretical spectral functions, since $k_BT\ll 20$~meV. Instead, we believe that it is due to the presence of the gap arc.

Further support for this interpretation comes from Fig.~4 of Ref.~\cite{Yan23}, where the Fermi function has been removed from the data. For Cut~2 close to the antinode the authors observe a feature at $E\approx -20$~meV at $T=17$~K and 50~K, but not at 70~K (close to $T_c = 73$~K of the studied sample) and at higher temperatures.  Our Fig.~\ref{fig:antinodal_yan} clearly demonstrates, for reasonable values of the MRDP parameters, that the feature at $E$ may indeed be interpreted as a gap arc. Moreover, comparison between panels (a) and (b) of Fig.~\ref{fig:antinodal_yan} illustrates the extreme sensitivity of the second-derivative technique developed in Ref.~\cite{Yan23} to the presence of the gap arcs, which are barely visible in panel (a), but very prominent in panel (b).

As already explained, gap arcs should be most easily observable if $\Gamma_s\gg\Delta\gtrsim\Gamma$. In order to achieve this situation, the off-plane disorder should be as large as possible, while the CuO$_2$ planes should be as clean as possible. In Ref.~\cite{Yan23}, where the gap arcs might already have been observed, large off-plane disorder of the overdoped Bi2212 samples was presumably achieved by excess oxygen in the BiO planes. Alternatively, substitutions of Pb for Bi or of Y for Ca can be considered. Although such substitutions change the electron count, techniques exist for a precise control of the doping level in their presence \cite{Hobou09}.

\section{Comparison of MRDP to previous work}\label{sec:Comparison} 
\subsection{Treatment of impurities}
The study of the effect of impurities on anisotropic superconductors has a long history. Even before the discovery of the cuprates, the subject became popular in the context of heavy fermion superconductivity. The standard theoretical tool for treating this problem is the self-consistent T-matrix approach. Recently, this method received a renewed interest within the so-called dirty $d$-wave theory of overdoped cuprates, see \cite{LeeHone18} and references therein. Several types of impurities characterized by different phase shifts have been treated by this method, with particular emphasis on impurities in the Born and unitary limits. It was assumed that the electrons scatter on the impurities isotropically, and in this sense these are particular cases of large-angle scattering in our language.   Within the dirty $d$-wave theory, for isotropic scattering the off-diagonal self-energy $\Phi({\bf k},\omega)$ is simply equal to the gap $\Delta_{\bf k}$ and only the diagonal self-energy $\Sigma(\omega)$ exhibits a non-trivial frequency dependence. The function $\Sigma(\omega)$ has to be determined from a self-consistent equation, which can be solved only numerically. Within MRDP, we instead model the large-angle impurity scattering analytically, by an appropriately chosen parameter $\Gamma$. (Of course, a small part of the scattering is of the forward-scattering type even for isotropically scattering impurities.)

The Abrahams-Varma proposal that a large part of impurity scattering has to be of the forward-scattering type \cite{Abrahams00} has been studied within the self-consistent Born approximation (SCBA) in several later works. In particular, in Refs.~\cite{Markiewicz04,Zhu04} the authors found that in presence of dominant forward scattering, the electron spectral function is given by the same formula as our Eq.~\eqref{eq:spectral_no_gamma} valid in the special case when  $\Gamma=0$. This provides further support for MRDP. Note, however, that within our approach we did not have to assume that the scattering is weak. 

As noted in Ref.~\cite{Abrahams00}, within SCBA the forward-scattering rate scales with the local ($\theta$-dependent) density of states, and therefore grows towards the antinode. Explicit calculations  within SCBA show that a qualitatively similar angular dependence applies also in the superconducting state \cite{Graser08}. It is pleasing to note that, as can be seen from Fig.~\ref{fig:uhly}, from the MRDP fits to experimental data  one finds that $\Gamma_s(\theta)$ is a growing function of $\theta$. It is worth pointing out that, within MRDP, there exists yet another argument why $\Gamma_s$ should increase towards the antinode: as explained in Appendix~\ref{appendix_heuristics}, the momentum cutoff $q_c(\theta)$ which controls the magnitude of $\Gamma_s$ grows with increasing $\theta$.

To summarize, our full expression for the Green's function Eq.~\eqref{eq:green_f} provides a simple analytical formula where both types of scattering, large-angle as well as small-angle, are treated on the same footing. Therefore it can be thought of as an effective description of the combined effect of in-plane \cite{LeeHone18} and out-of-plane \cite{Markiewicz04,Zhu04} impurities studied earlier.

\subsection{Comparison with the Norman formula}
To start with, let us point out that, within the standard Eliashberg theory of the superconducting state, one introduces two self-energies, $\Sigma(\omega)$ and $\Phi(\omega)$, both of which depend on frequency $\omega$ and possibly also on the Fermi-surface position ${\bf k}$. In terms of these functions, the full Nambu-Gorkov $2\times 2$ Green's function reads
\begin{equation}
{\hat G}({\bf k},\omega)=
\frac{(\omega-\Sigma)\tau_0+\Phi\tau_1+\varepsilon_{\bf k}\tau_3}
{(\omega-\Sigma)^2-\Phi^2-\varepsilon_{\bf k}^2},
\label{eq:green_matrix}
\end{equation}
where $\tau_i$ are the Pauli matrices. The Green's function Eq.~\eqref{eq:green_f} is its 11 component, $G({\bf k},\omega)\equiv {\hat G}_{11}({\bf k},\omega)$. In the ARPES literature, the Green's function is often described by $G({\bf k},\omega)=[\omega-\varepsilon_{\bf k}-{\cal S}({\bf k},\omega)]^{-1}$, where ${\cal S}({\bf k},\omega)$ is what we will call 
the "photoemission self-energy". The Eliashberg theory implies that
\begin{equation}
{\cal S}({\bf k},\omega)=
\Sigma+\frac{\Phi^2}{\omega-\Sigma+\varepsilon_{\bf k}}.
\label{eq:photoemission_self}
\end{equation}

In the widely used Norman's formula (NF) for the photoemission self-energy \cite{Norman98} one assumes that $\Phi=\Delta$, and replaces $\Sigma$ by $-i\Gamma_1$ in the first term of Eq.~\eqref{eq:photoemission_self}, and by $-i\Gamma_0$ in the denominator of the second term. Thus, the NF Green's function is not compatible with the Eliashberg theory. 

\begin{figure}[t]
\includegraphics[width = 8.3 cm]{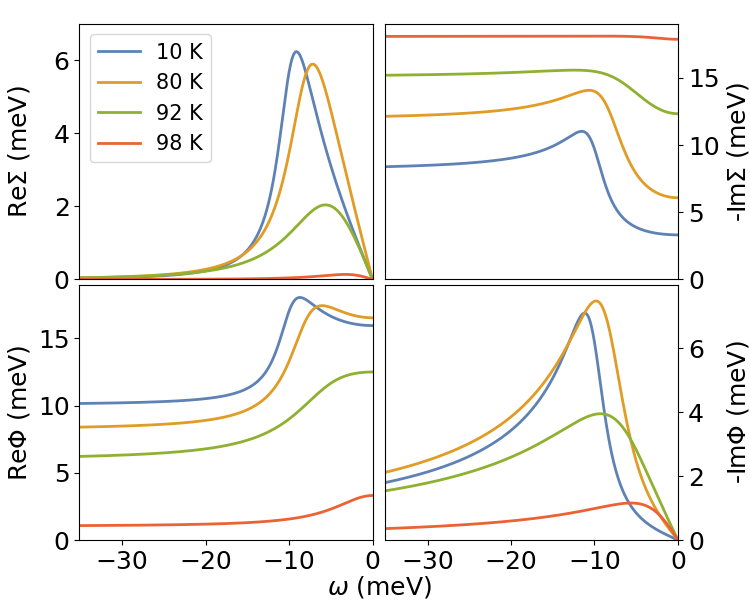}
\caption{Temperature dependence of the normal and anomalous self-energies $\Sigma(\omega)$ and $\Phi(\omega)$, respectively,  calculated using Eq.~\eqref{eq:self_energies} and the MRDP parameters from Fig.~\ref{fig:leporelo2} at $\theta=10^\circ$. At $T=98$~K, from MRDP fits we find $\Delta = 1$~meV, $\Gamma = 5.5$~meV, and $\Gamma_s  = 12.6$~meV.}
\label{fig:selfenergy}
\end{figure}

In \cite{Herman17a} we have shown that, for $\varepsilon_{\bf k}\neq 0$, NF is inconsistent with the exact sum rules for the particle number, because the electron-like and hole-like branches exhibit different scattering rates. A particularly useful way to see this is as follows. Let us introduce auxiliary scattering rates $\Gamma_+=(\Gamma_1+\Gamma_0)/2$ and $\Gamma_-=(\Gamma_1-\Gamma_0)/2$ and let us define $\Omega_{\rm NF}(\omega)$, $U_{\rm NF}^2(\omega)$ and $V_{\rm NF}^2(\omega)$ exactly as in Section~II, just replacing $\Gamma$ by $\Gamma_+$. Then the NF Green's function can be written very similarly as Eq.~\eqref{eq:green_f}:
\begin{equation}
 G_{\rm NF}({\bf k},\omega)=
\frac{U_{\rm NF}^2(\omega)}{\Omega_{\rm NF}+i\Gamma_{-}-\varepsilon_{\bf k}}
+\frac{V_{\rm NF}^2(\omega)}{\Omega_{\rm NF}-i\Gamma_{-}+\varepsilon_{\bf k}}.
\label{eq:norman}
\end{equation}
The different signs in front of $\Gamma_-$ in the two terms clearly show that the scattering rates of  electron-like and hole-like branches are different. It is worth pointing out that the Dynes Green's function Eq.~\eqref{eq:green_f} can be understood as the simplest particle-hole symmetry-restoring correction to Norman's formula.

\begin{figure*}[t!]
\includegraphics[width = 15.2 cm]{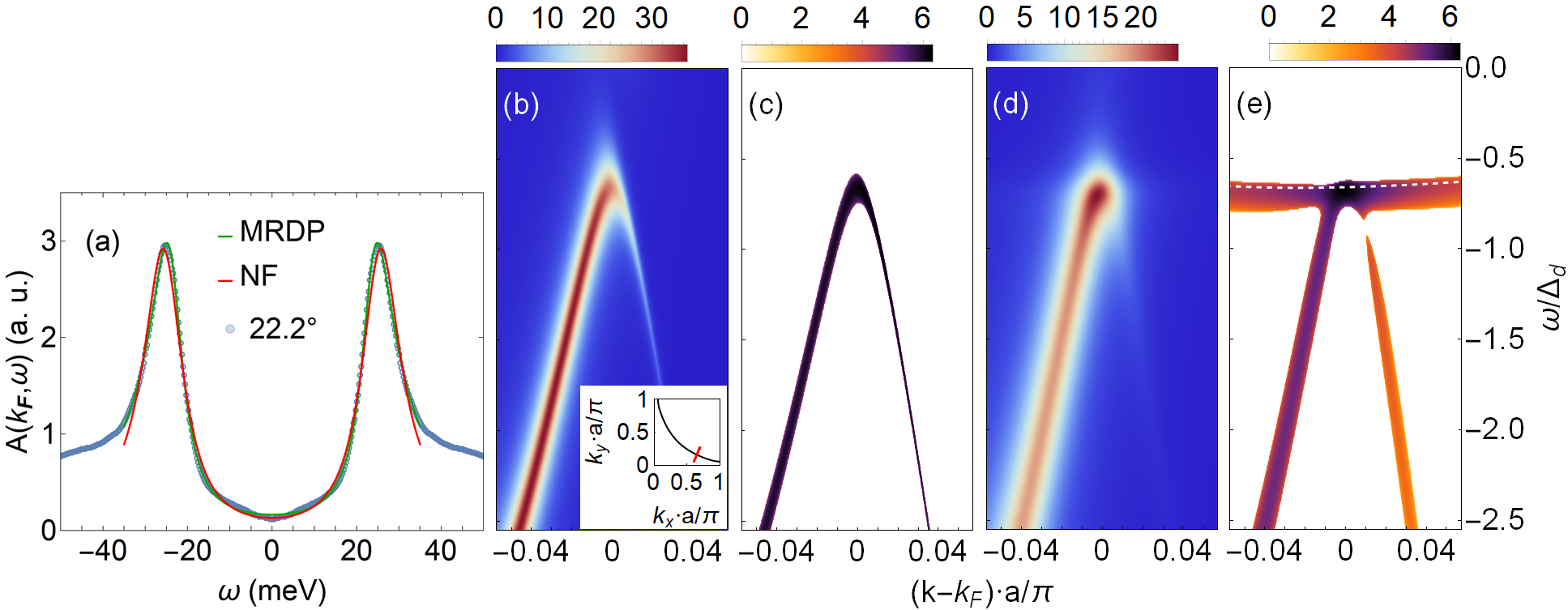}
\caption{(a) Fermi-surface spectral function at angle $\theta=22.2^\circ$ for optimally doped Bi2212 at $T=10$~K taken from Ref.~\cite{Kondo15} and its fits by MRDP and NF, see text. (b,d): Tomographic maps of $A(k,\omega)$. (c,e): Logarithm of the second derivatives with respect to energy $\log[-A''(k,\omega)]$ of the same data. Only points where $A''(k,\omega)<0$ are shown.  (b,c): Prediction of NF. (d,e): Prediction of MRDP. White dashed line in (e): the function $\omega=-\Delta_{\bf k}$ along the tomographic cut. In (b-e), the parameters entering the formulas for $A(k,\omega)$ were determined from the fits in (a).}
\label{fig:norman}
\end{figure*}

Further problems with NF are the following. Although NF does reproduce the Dynes formula for TDoS, it predicts that the Dynes parameter in Eq.~\eqref{eq:Dynes_formula} is $\Gamma_+$. This means that $\Gamma_1$ and $\Gamma_0$ cannot be interpreted as single-particle and pair-scattering rates, respectively, as suggested in Ref.~\cite{Kondo15}. Moreover, NF also cannot explain the large difference of scattering rates between ARPES and tunneling, since a small value of $\Gamma_+$ requires that both, $\Gamma_{0}$ and $\Gamma_{1}$, have to be small. But in such case $\Gamma_-$ cannot be large.

On the other hand, MRDP is perfectly consistent with the Eliashberg theory. In case of Dynes superconductors, the self-energies defined in Eq.~\eqref{eq:green_matrix} are given by
\begin{equation}
\Sigma(\omega)=-i\Gamma-i\Gamma_s\frac{\omega+i\Gamma}{\Omega},
\quad
\Phi(\omega)=\left(1+\frac{i\Gamma_s}{\Omega}\right)\Delta.
\label{eq:self_energies}
\end{equation}
As an example, making use of the MRDP parameters from Fig.~\ref{fig:leporelo2} at $\theta=10^\circ$, we have calculated the $T$-dependence of the self-energies using Eq.~\eqref{eq:self_energies}. The results are shown in Fig.~\ref{fig:selfenergy}. The qualitative agreement in the low-energy region with the findings of Ref.~\cite{Bok16}
is obvious: with decreasing $T$, the peak of ${\rm Re}\Sigma(\omega)$ grows at the gap position, while the contribution of $\Gamma_s$ to $-{\rm Im}\Sigma(\omega)$ is strongly suppressed below the gap. Of course, the predictions of MRDP differ from the experimental results at large frequencies. In fact, within MRDP $-{\rm Im}\Sigma(\omega)$ saturates to $\Gamma_s+\Gamma$ and ${\rm Re}\Sigma(\omega)$ vanishes, while the actual experimental result is that $-{\rm Im}\Sigma(\omega)$ grows with $|\omega|$ and ${\rm Re}\Sigma(\omega)$ exhibits a finite background. Also the gap-related features of $\Phi(\omega)$ in the low-energy region are seen to be qualitatively consistent with Ref.~\cite{Bok16}.

Finally we make an additional observation  that, even if we take NF as a purely phenomenological formula for $\omega<0$, it does not correctly describe the observed low-temperature data. In order to compare the predictions of MRDP and NF away from the Fermi surface, we have first determined the parameters for both theories from the Fermi-surface data. To this end, we analyzed the data at $\theta=22.2^\circ$ and $T=10$~K in optimally doped Bi2212 from Ref.~\cite{Kondo15}. From the fit to NF we find $\Gamma_0\approx 2.7$~meV, $\Gamma_1\approx 8.6$~meV, and $|\Delta|\approx 26$~meV. On the other hand, from MRDP fits we find $\Gamma\approx 3.7$~meV, $\Gamma_s\approx 14$~meV, and $|\Delta|\approx 24$~meV. As shown in Fig.~\ref{fig:norman}, the quality of both fits is comparable.

Afterwards we have calculated the tomographic maps predicted by both theories for the same angle $\theta=22.2^\circ$, making use of the parameters found at the Fermi surface. The results of such analysis are plotted in Fig.~\ref{fig:norman}. Note that only MRDP predicts the existence of the bright spot at the Fermi surface and no such enhancement is predicted by NF.

\section{Conclusions}\label{sec:Conclusion}
In conclusion, we have shown that ARPES spectroscopy of the superconducting state offers a unique possibility to distinguish between forward- and large-angle scattering processes in the cuprates. Although we have considered only strictly elastic scattering, we have argued that our conclusions apply at least qualitatively also to inelastic scattering by low-lying modes. A detailed study of the angle- and temperature-dependence of the forward- and large-angle scattering rates $\Gamma_s$ and $\Gamma$, respectively, might therefore provide hitherto inaccessible information on the dynamics of the electrons in the cuprates \cite{note}. 

We find that $\Gamma_s$ is strongly anisotropic and much larger than $\Gamma$, which is only weakly angle-dependent. The difference in magnitudes of the two scattering rates explains the well-known difference of the smearings observed by ARPES (which is essentially $\Gamma_s+\Gamma$) and STM (which is $\Gamma$). We believe that this difference is a real phenomenon and not an experimental artifact caused by lower resolution of the ARPES experiments. The reason is threefold. First, the magnitude of $\Gamma$ extracted from the tomographic density of states (i.e., from ARPES) is very similar to that obtained from STM. Second, the energy resolution of recent ARPES experiments is typically 1~meV \cite{Yan23}, much less than the observed scattering rates $\Gamma_s$. Third, it might be challenging to explain the extracted angle-dependent scattering rate $\Gamma_s$ as an experimental artifact.

We have also suggested a novel method for measuring the superconducting gap $\Delta_{\bf k}$ away from the Fermi surface by observation of the so-called gap arcs. The method should work best at low $T$, when pair-breaking scattering is minimized. Large forward scattering is also needed, and this can be realized by selectively increasing disorder outside the CuO$_2$ planes and/or by studying the antinodal region. We have argued that gap arcs may have already been seen recently \cite{Yan23}.

From a more general perspective, the observation that MRDP works in the context of the cuprates is not that surprising. In fact: the relation between the Dynes phenomenology and the BCS model is the same as the relation between a metal with a finite electron lifetime and the model of a free Fermi gas. The only difference has to do with the well known fact that a superconductor is described by two lifetimes: pair-conserving and pair-breaking. Therefore we believe that Dynes phenomenology should be the model of first choice when analyzing ARPES data in any real superconductor.

\begin{acknowledgments} 
This work was supported by the Slovak Research and Development Agency under Contract No.~APVV-23-0515.
\end{acknowledgments}

\appendix
\section{Heuristic justification of MRDP}
\label{appendix_heuristics}
The Dynes phenomenology for $s$-wave superconductors can be justified using the coherent-potential approximation (CPA, for a review, see \cite{Elliott74}) for the Green's function of an electron in a disordered superconductor. In order to explain the main idea of CPA, let us start by defining the disorder-averaged frequency-dependent local $2\times 2$ Nambu-Gorkov Green's function $\mathcal{\hat{G}}_n={\mathcal N}^{-1}\sum_{\bf k}\hat{G}({\bf k},i\omega_n)$, where ${\mathcal N}$ is the number of lattice sites and $\omega_n$ is the Matsubara frequency. Within CPA one requires that $\mathcal{\hat{G}}_n$ satisfies the equation \cite{Herman17a}
\begin{equation}
\mathcal{\hat{G}}_n=\left\langle\left(
\mathcal{\hat{G}}_n^{-1}-\hat{W}+\hat{\Sigma}_n
\right)^{-1}\right\rangle,
\label{eq:CPA}
\end{equation} 
where $\hat{W}$ describes the fluctuating local potential at the studied lattice site and $\hat{\Sigma}_n$ is the disorder-averaged electron self-energy. The  angular brackets denote averaging with respect to disorder. In Ref.~\cite{Herman16} it was assumed that, in $s$-wave superconductors, the potential is given by ${\hat W}=\Delta \tau_1+U\tau_3 +V\tau_0$, where $\Delta$ is the (spatially constant) gap in absence of impurities, $U$ is the random non-magnetic potential, $V$ is the random magnetic field, and  $\tau_i$ are the Pauli matrices in the Nambu-Gorkov space. The distribution functions of $U$ and $V$, $P_U(U)$ and $P_V(V)$, satisfy the normalization conditions $\int dU P_U(U)=\int dV P_V(V)=1$. 

\begin{figure*}[]
\includegraphics[width = 15.2 cm]{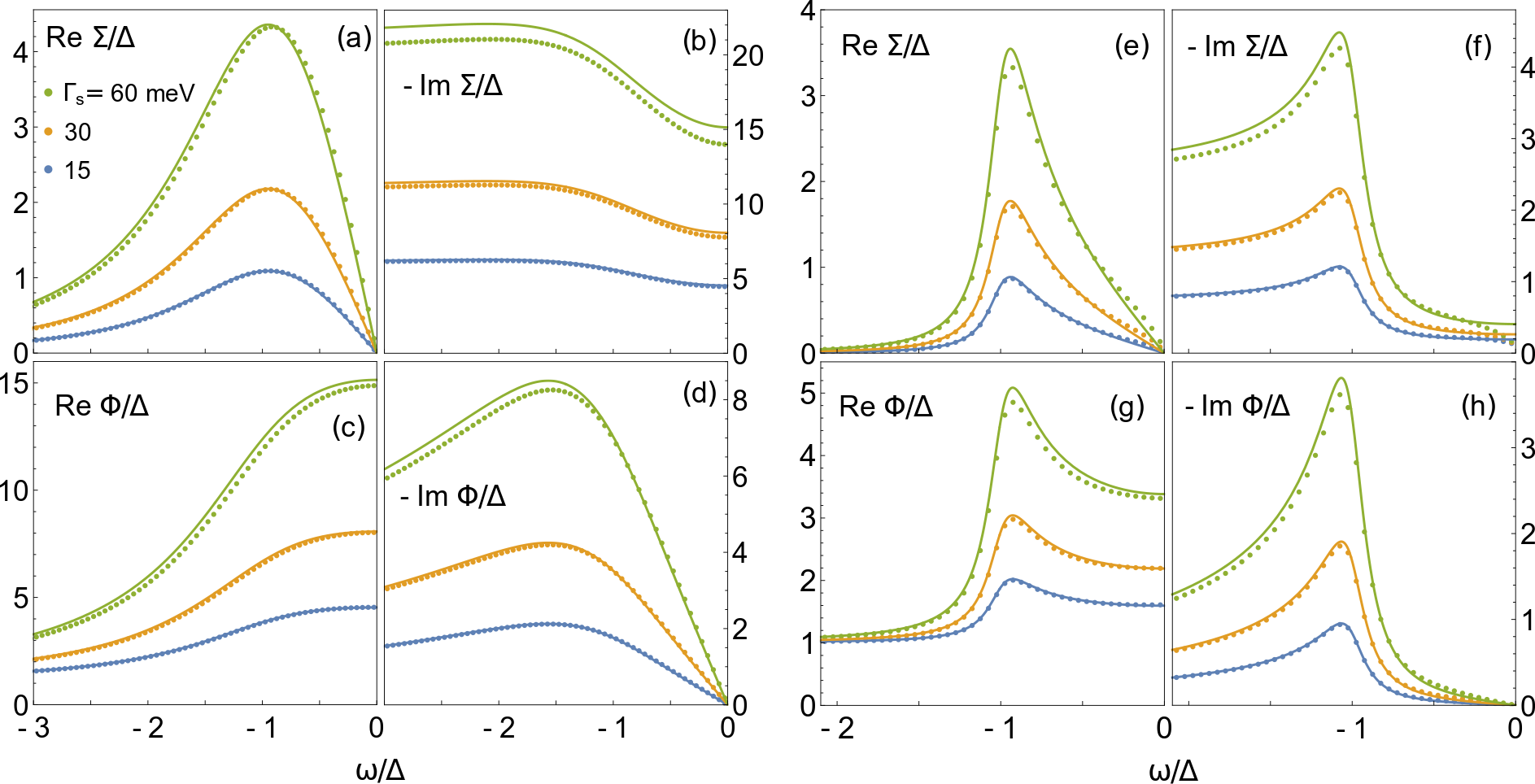}
\caption{Diagonal and off-diagonal self-energies $\Sigma(\omega)$ and $\Phi(\omega)$ calculated using Eq.~\eqref{eq:auxiliary_z}, taking for the distribution $P_V(V)$ Lorentzians of width $\Gamma$  and for $P_U(U)$ Gaussians with width $U_0$ (data points). Furthermore we assume that $N_0=1/(8t)$. The results are compared with predictions of the Dynes phenomenology with parameters $\Gamma$ and $\Gamma_s=\pi N_0 U_0^2$ (solid lines). (a-d): $\Delta=3$~meV and $\Gamma=3$~meV, roughly corresponding to near-nodal data. (e-h): $\Delta=25$~meV and $\Gamma=3$~meV, relevant away from the node. In every panel, curves from bottom to top correspond to $\Gamma_s=15$~meV, 30~meV, and 60~meV.}
\label{fig:CPA}
\end{figure*}

We would like to point out that the more commonly used approaches to dirty superconductors, for instance discussions of the Born or unitary-scattering limits on dilute impurities with concentration $x$, can be treated within CPA by considering special choices of the distribution functions such as $P_U(U)=(1-x)\delta(U)+x\left[\delta(U-U_0)+\delta(U+U_0)\right]/2$. Compared with these more conventional approaches, within our CPA-based theory we can treat a much wider variety of physical situations, in particular also cases with dense distributions of impurities.

Note that Eq.~\eqref{eq:CPA} has a natural interpretation: the self-energy $\hat{\Sigma}_n$ has to be chosen so that, on average, it compensates for the presence of the random potential $\hat{W}$. In MRDP we postulate that, if the local Green's function $\mathcal{\hat{G}}_n$ is replaced by the tomographic local Green's function $\mathcal{\hat{G}}_n(\theta)=\int dk \hat{G}(k,\theta,i\omega_n)$, the physically well motivated Eq.~\eqref{eq:CPA} remains locally valid on a given cut $\theta$ also in the anisotropic case. 

Only non-magnetic scattering is considered in MRDP. Let us assume that the random potential in a given CuO$_2$ plane is $u({\bf r})$ and its Fourier transform, causing scattering with momentum transfer ${\bf q}$, is $u_{\bf q}$. Disorder in an ensemble of such planes is described by the probability distribution ${\cal P}(\{u_{\bf q}\})$. As a result of impurity scattering the electron experiences, in general, a change of the gap $\delta \Delta$. Let us define for every angle $\theta$ a critical momentum transfer $q_c(\theta)$ by requiring that for all momentum transfers $q<q_c$ the scattering essentially does not change the gap, $|\delta \Delta|\ll |\Delta|$. Obviously, $q_c$ will be momentum dependent: it vanishes at the node and grows towards the antinode. In what follows, scattering with momentum transfer $q<q_c$ will be called forward scattering. As is well known \cite{Millis88}, such scattering is not pair breaking and therefore it can be described by a simple potential scattering term $U\tau_3$ with $U\sim \sum_{q<q_c} u_{\bf q}$, precisely as in the s-wave case. Within this approach, the actual type of impurities in the system is encoded in the distribution function $P_U(U)$, which may be $\theta$-dependent.

On the other hand, "large-angle" scattering with $q>q_c$ does appreciably change the gap value. Within MRDP we model it by scattering-induced fluctuations of the gap described by the term $V\tau_1$, where $V$ is the random deviation of the gap from its value at the studied tomographic cut $\theta$. Note that, again, the distribution function $P_V(V)$ depends on $\theta$, as well as on the detailed properties of large-angle scattering, i.e. on the distribution of $u_{\bf q}$ at $q>q_c$. To summarize, in MRDP we take $\hat{W}=[\Delta(\theta)+V] \tau_1+U\tau_3$, where $U$ and $V$ are fluctuating fields. 

In what follows we assume that the self-energy for angle $\theta$ can be parametrized by $\hat{\Sigma}_n(\theta)=-i\Gamma_n(\theta)\tau_0+\Phi_n(\theta)\tau_1$, as usual. From now on, for the sake of brevity we do not explicitly show the dependence on the chosen cut $\theta$. Similarly as in the $s$-wave case studied in Ref.~\cite{Herman16}, in the present case the CPA equation~\eqref{eq:CPA} can be rewritten as 
\begin{equation}
z_n=\left\langle\frac{w_n+\lambda}
{(w_n+\lambda)(w_n^\ast+\lambda)+\mu^2}\right\rangle_{\lambda,\mu},
\label{eq:auxiliary_z}
\end{equation} 
where instead of $U$ and $V$ we have introduced dimensionless fluctuating fields $\mu=\pi N_0 U$ and $\lambda=\pi N_0V$, with $N_0$ the (tomographic) density of states in the normal state. When deriving Eq.~\eqref{eq:auxiliary_z}, we have assumed that the distribution function $P_U(U)$ is even, $P_U(-U)=P_U(U)$. The auxiliary complex numbers $z_n$ and $w_n$ are functions of the diagonal and off-diagonal self-energies $\Gamma_n$ and $\Phi_n$:
\begin{eqnarray}
z_n&=&\frac{\Phi_n+i(\omega_n+\Phi_n)}{\sqrt{(\omega_n+\Gamma_n)^2+\Phi_n^2}},
\nonumber
\\
w_n&=&z_n+\pi N_0(\Delta-\Phi_n-i\Gamma_n).
\label{eq:definition_zn}
\end{eqnarray} 
The solutions of the coupled Eqs.~\eqref{eq:auxiliary_z},~\eqref{eq:definition_zn} for the self-energies $\Gamma_n$, $\Phi_n$ obviously depend on the precise form of the distribution functions $P_U(U)$ and $P_V(V)$. Note that Eq.~\eqref{eq:auxiliary_z} is very similar to Eq.~(5) derived for the $s$-wave case in Ref.~\cite{Herman16}, while the definitions Eqs.~\eqref{eq:definition_zn} are in fact identical to those of Ref.~\cite{Herman16}. 

One checks readily that, in absence of forward scattering, i.e. for $\mu=0$, the present set of equations is in fact identical to the equations studied in Ref.~\cite{Herman16} (also for $\mu=0$). From here it follows that, in the special case when $\mu=0$ and if the distribution function $P_V(V)$ is described by a Lorentzian of width $\Gamma$, the tomographic Green's function is given by the Dynes formula Eq.~\eqref{eq:green_f} with parameters $\Delta$, $\Gamma$, and $\Gamma_s=0$. For more realistic distribution functions $P_V(V)$ we do not know the solution analytically, but if $P_V(V)$ is sufficiently broad, we expect that the Green's function might still be reasonably well approximated by Eq.~\eqref{eq:green_f}. 

By fitting the experimental data we have argued in the main text that forward scattering is definitely present in the cuprates. Unfortunately, in that case Eq.~\eqref{eq:auxiliary_z} can not be solved analytically even if $P_V(V)$ is described by a Lorentzian. That is why in this case we have solved Eq.~\eqref{eq:auxiliary_z}  numerically. For the distribution functions $P_U(U)$ we took Gaussians centered at $U=0$ with several widths $U_0$ leading to experimentally relevant values of $\Gamma_s$. Figure~\ref{fig:CPA} shows that the numerically obtained self-energies $\Sigma(\omega)$ and $\Phi(\omega)$ are very well described by the Dynes phenomenology not only for $\Gamma_s=0$, but up to $\Gamma_s=60$~meV, larger than required to fit the experiment.

We would like to point out that the gap $\Delta(\theta)$ enters our theory only as a phenomenological input. In order to derive a truly microscopic theory for the gap, one would need to take into account self-consistency between the distribution function $P_V(V)$ and the angular variation of $\Delta(\theta)$. This task is beyond the scope of the present paper.

\section{Spectral functions at the Fermi surface} 
\label{appendix_fermi_surface}
\begin{figure*}[b!]
\includegraphics[width = 15.5 cm]{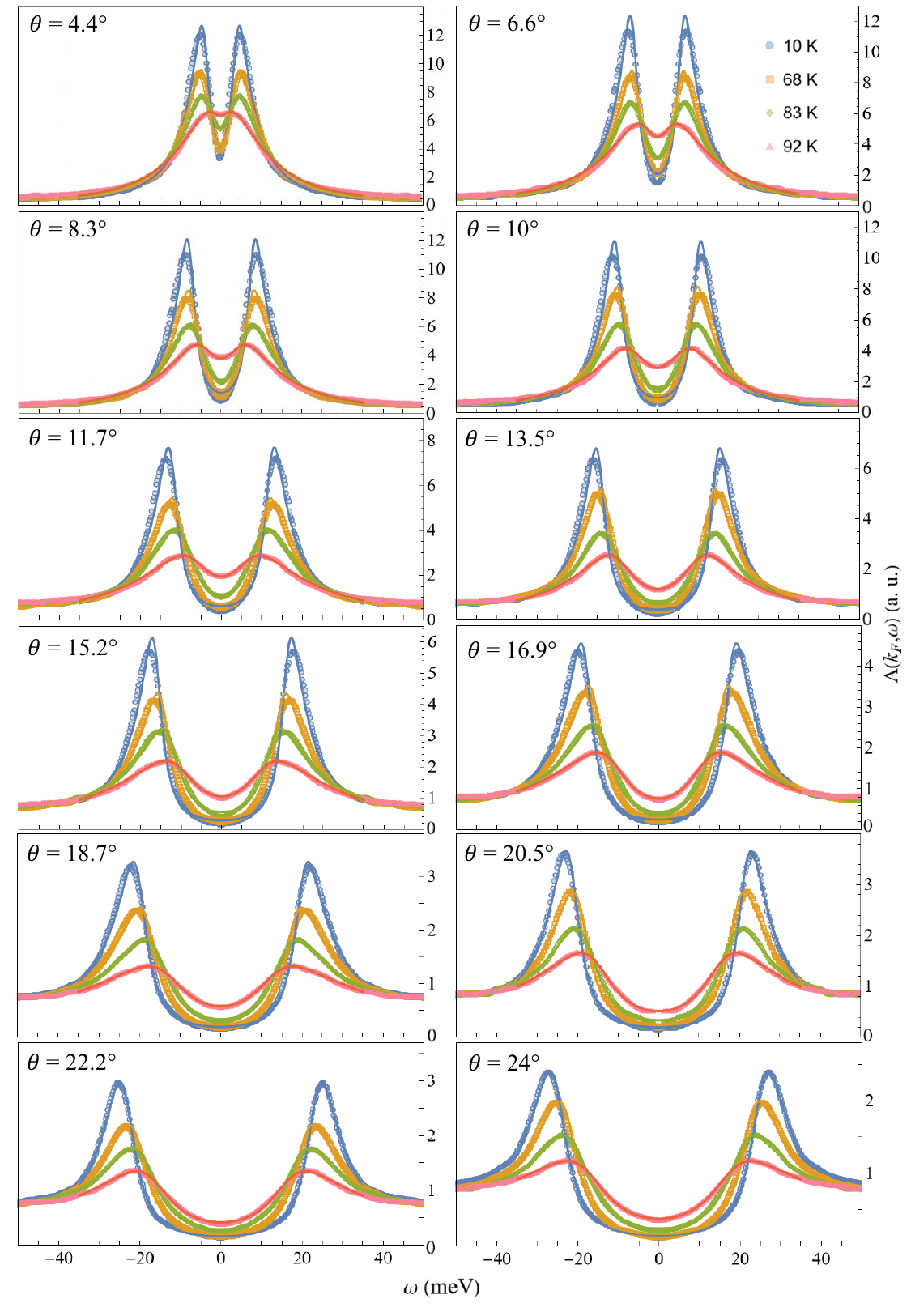}
\caption{Spectral functions at $k=k_F$ in optimally doped Bi2212 taken from Ref.~\cite{Kondo15} (symbols), and their MRDP fits at energies $|\omega|<35$~meV (lines) for selected temperatures.}
\label{fig:leporelo1}
\end{figure*}

\begin{figure*}[t!]
\includegraphics[width = 15.5 cm]{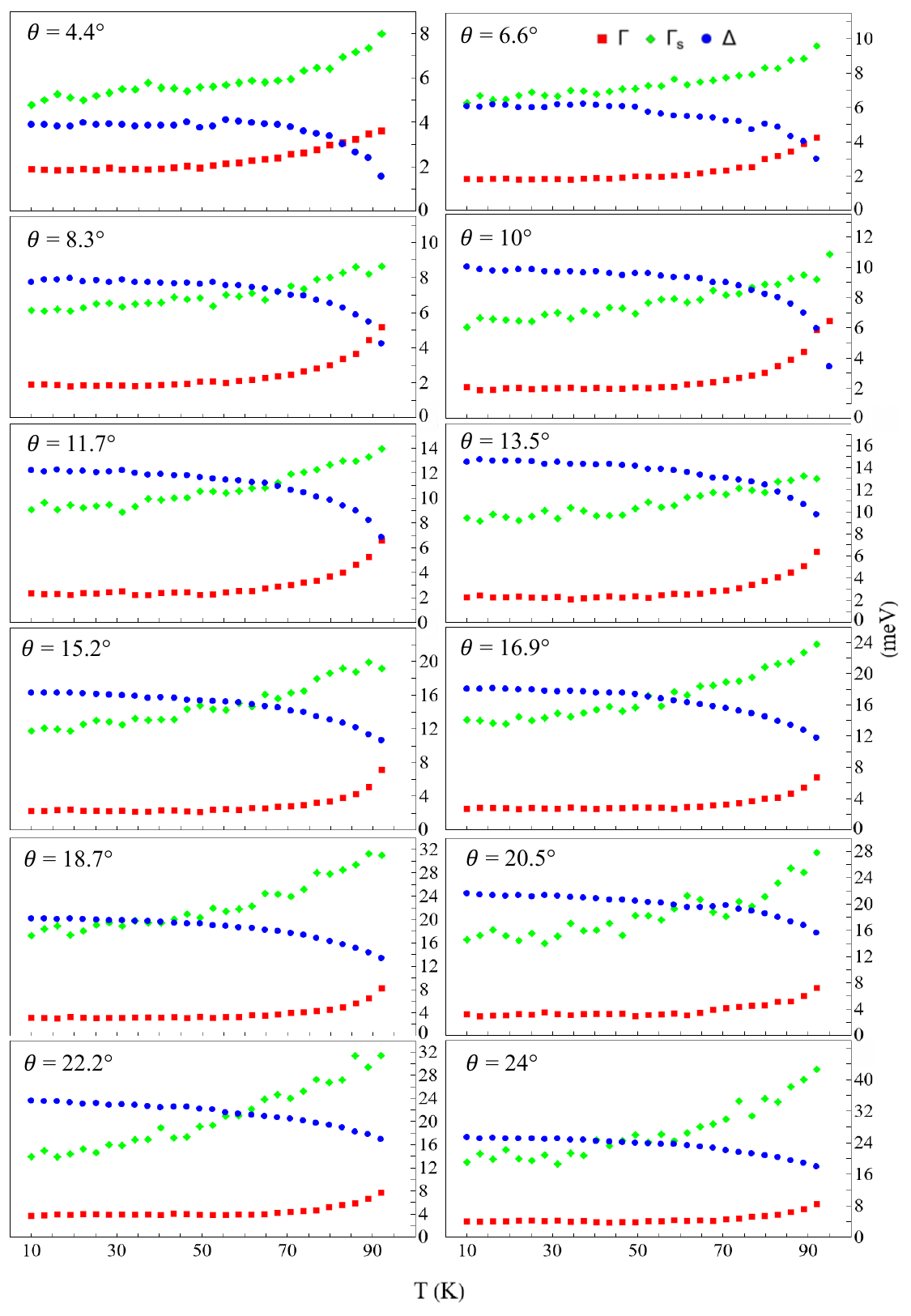}
\caption{$T$-dependence of the parameters $\Gamma$, $\Gamma_s$, and $\Delta$ determined by MRDP fits of the spectral functions at $k=k_F$ in optimally doped Bi2212 taken from Ref.~\cite{Kondo15}.}
\label{fig:leporelo2}
\end{figure*}

In Figs.~\ref{fig:leporelo1},~\ref{fig:leporelo2} we show the results of the MRDP analysis for all angles reported in Ref.~\cite{Kondo15}. As one can see, the overall quality of the fits is good and it improves with increasing $\theta$. The $T$-dependence of the extracted parameters is  for all angles similar to the results for $\theta=24^\circ$ reported in the main text.

\clearpage

\begin{figure}[!t]
\includegraphics[width = 8.7 cm]{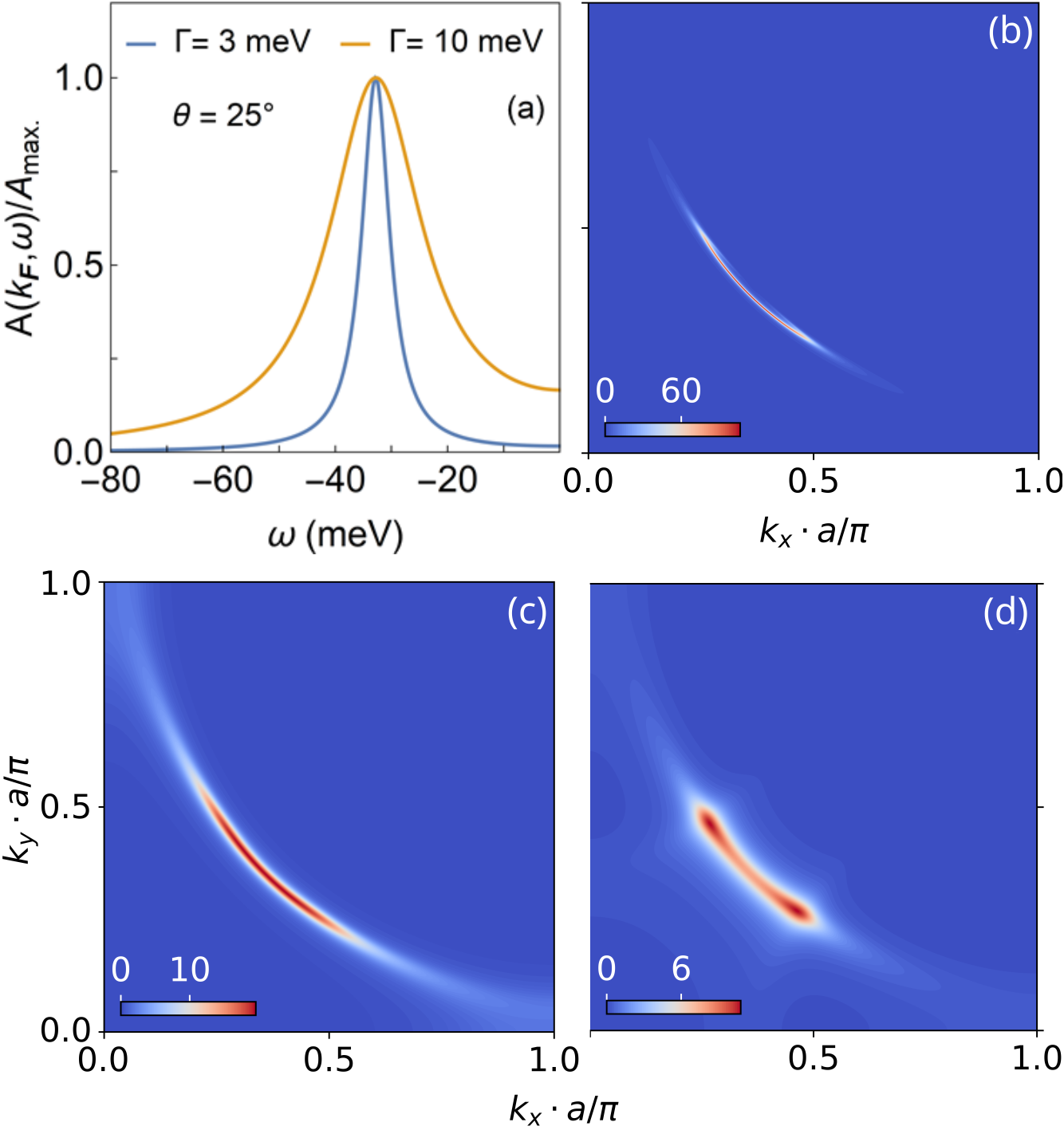}
\caption{Spectral functions for $\Delta_d= 45$~meV and momentum-independent scattering rates $\Gamma$ and $\Gamma_s$. (a): Energy distribution curves at the Fermi surface point $\theta=25^\circ$ for two values of $\Gamma$, assuming that $\Gamma_s=0$. (b-d): Momentum maps of $A({\bf k},\omega)$ at energy $\omega=-20$~meV. (b): Hypothetical case $\Gamma=3$~meV and $\Gamma_s=0$. (c,d): Momentum maps taking into account the finite momentum resolution $\delta k\approx 0.0055$~\AA$^{-1}$ reported in \cite{Shah25}. (c): Experimentally relevant case $\Gamma=10$~meV and $\Gamma_s=0$. (d): Prediction of MRDP with $\Gamma = 3$~meV and $\Gamma_s = 35$~meV.  Note that in (b,c) the spectral function is largest in the middle of the banana, while in (d) the bright spots are clearly visible.}
\label{fig:momentum_map_suppl}
\end{figure}

\section
{\bf Bright spots in momentum maps} 
\label{appendix_momentum maps}

In Fig.~\ref{fig:momentum_maps}c from the main text we plot the momentum map for $\Gamma=10$~meV and $\Gamma_s=0$. Our choice of the pair-breaking rate $\Gamma$ is explained in Fig.~\ref{fig:momentum_map_suppl}a, where we plot the energy distribution curves relevant for the experiment Ref.~\cite{Shah25} for two different values of $\Gamma$, assuming that no forward scattering is present. Comparison with the experimental Fig.~1c in Ref.~\cite{Shah25} shows that, for instance at the Fermi surface point for $\theta=25^\circ$, $\Gamma=10$~meV is the minimal value compatible with their data, because smaller values of $\Gamma$ correspond to very narrow spectral functions. Since in our Fig.~\ref{fig:momentum_maps}c from the main text there are no bright spots at the ends of the bananas, we conclude that experimental observation of bright spots in \cite{Shah25} requires that a finite $\Gamma_s$ must be present. 

One might nevertheless ask whether, still in the absence of $\Gamma_s$ but for much smaller values of $\Gamma$, the bright spots do appear. One can show analytically that this is not the case. To this end, let us compare the spectral weight for $\Gamma_s=0$ at a fixed energy $\omega<0$ in two points of the equienergetic line $E_{\bf k}=-\omega$: in the nodal direction and at the tip of the banana. Making use of Eq.~\eqref{eq:spectral_no_gammas} we find that $A_{\rm node}=1/(\pi\Gamma)$ and $A_{\rm tip}=1/(2\pi\Gamma)$, thus $A_{\rm tip}$ is always smaller than $A_{\rm node}$, regardless of the magnitude of $\Gamma$. For an explicit example with a hypothetical small $\Gamma$, see Fig.~\ref{fig:momentum_map_suppl}b. 

ARPES experiments have a finite resolution in momentum space $\delta k$. Let us denote the length of the banana as $k_1$, the Fermi velicity as $v_F$, and the gap-function slope as $v_\Delta$. In the limit $\delta k/k_1 \ll (v_\Delta/v_F)^2$ one can simply estimate the observed spectral functions $\widetilde{A}_{\rm node}$ and $\widetilde{A}_{\rm tip}$, which take into account also the finite value of momentum resolution. By taking convolutions with Lorentzians, one finds
\begin{eqnarray}
\widetilde{A}_{\rm node}&=&\frac{1}{\pi v_F \delta k}
\arctan\left(\frac{v_F\delta k}{\Gamma}\right),
\nonumber\\
\widetilde{A}_{\rm tip}&=&\frac{1}{2\pi v_\Delta \delta k}
\arctan\left(\frac{v_\Delta \delta k}{\Gamma}\right).
\label{eq:convolutions}
\end{eqnarray}
In the limit of very small $\Gamma$, bright spots would be present at the tips of the bananas even within the single-$\Gamma$ model, since $v_\Delta\ll v_F$ \cite{Vishik10}. However, the experimentally relevant values of $\Gamma\gtrsim 10$~meV and the momentum resolution $\delta k\approx 0.0055$~\AA$^{-1}$ taken from \cite{Shah25}, together with $v_F\approx 1.5$~eV\AA\: \cite{Vishik10}, imply that $v_F\delta k/\Gamma\lesssim 0.83$. Therefore $\widetilde{A}_{\rm node}\approx A_{\rm node}$ is twice as large as $\widetilde{A}_{\rm tip}\approx A_{\rm tip}$, and no bright spots are predicted by a model with $\Gamma_s=0$, in agreement with the main text. For an explicit numerical confirmation of this conclusion, see Fig.~\ref{fig:momentum_map_suppl}c. 

Finally, in Fig.~\ref{fig:momentum_map_suppl}d we demonstrate that, on the contrary, bright spots are clearly present within the MRDP description of the underdoped samples studied in \cite{Shah25}, if a reasonable set of parameters is used. A fully quantitative prediction of momentum maps within MRDP would require determination of the MRDP parameters from Fermi-surface data for the same sample. Unfortunately, such combined data for the same sample are currently not available and we urge the experimentalists to undertake such an analysis. 

\begin{figure*}[!t]
\includegraphics[width = 12.5 cm]{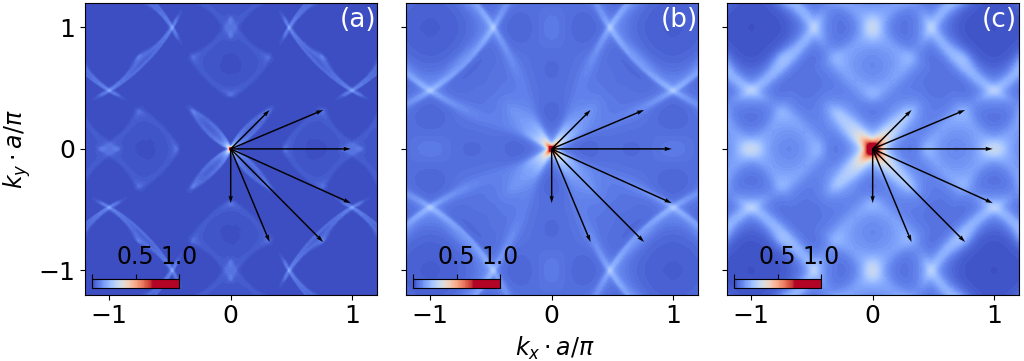}
\caption{Maps of the JDoS at $\omega=-20$~meV predicted by MRDP for a model superconductor with $\Delta_d=45$~meV and momentum-independent scattering rates $\Gamma$ and $\Gamma_s$, taking into account the finite momentum resolution $\delta k\approx 0.0055$~\AA$^{-1}$ reported in \cite{Shah25}. (a): $\Gamma_s=0$ and $\Gamma=1$~meV. (b): $\Gamma_s=0$ and $\Gamma=10$~meV. (c): $\Gamma_s=35$~meV and $\Gamma=3$~meV. The parameters in (b) and (c) are the same as those considered in Fig.~\ref{fig:momentum_map_suppl}c,d. The arrows denote the octet-model momenta connecting the ends of the four bananas in the full Brillouin zone.}
\label{fig:jdos}
\end{figure*}

Additional support for the finite value of $\Gamma_s$ in the MRDP spectral functions comes from the study of the joint density of states (JDoS) $S({\bf q},\omega)$, which is a convolution of the spectral functions at fixed $\omega$:
\begin{equation}
S({\bf q},\omega)=\sum_{\bf k} 
A({\bf k},\omega)A({\bf k}+{\bf q},\omega).
\label{eq:S_convolution}
\end{equation}
The authors of Ref.~\cite{Shah25} find that the experimental JDoS exhibits the same pattern as predicted by the so-called octet model \cite{McElroy03}, which was introduced to interpret the quasi-particle interference (QPI) observed by scanning tunneling spectroscopy of the cuprates. In Ref.~\cite{Shah25} it is therefore argued that JDoS can be understood as a proxy for the QPI spectra in the cuprates. We emphasize, however,  that it is not our ambition to interpret the QPI spectra here, and the present work deals exclusively with ARPES experiments.

Figure~\ref{fig:jdos} compares the patterns of JDoS in three situations. In panel (a) we consider a model without forward scattering and with a very small $\Gamma=1$~meV. One observes that the resulting pattern agrees with the octet model. However, if one requires that the model also explains the observed energy distribution curves \cite{Shah25}, a substantially larger $\Gamma=10$~meV is needed if we still require that $\Gamma_s=0$, see Fig.~\ref{fig:momentum_map_suppl}. Panel (b) shows that for such parameters the calculated JDoS is qualitatively different from experimental data. Finally, in panel (c) we show JDoS calculated from the full MRDP theory with realistic values of $\Gamma_s$ and $\Gamma$, see Fig.~\ref{fig:momentum_map_suppl}. 
Note that for this choice of parameters the momenta predicted by the octet model perfectly match the maxima of the theoretical JDoS, in agreement with the findings of Ref.~\cite{Shah25}.

\section
{\bf Bright spots in tomographic maps}
\label{appendix_tomographic_maps}
In Fig.~\ref{fig:tomographic_cut}a from the main text we plot the tomographic map of $A(k,\omega)$ for a largish value of $\Gamma=15$~meV and we find that in absence of forward scattering the bright spot at the Fermi surface, i.e. for $k=k_F$, does not appear. We would like to point out that this conclusion is valid for all values of $\Gamma$. The reason is the same as already pointed out: for $\Gamma_s=0$, at the Fermi surface we have $A_{\rm FS}=A_{\rm tip}=1/(2\pi \Gamma)$ and deep inside the Fermi surface $A_{\rm deep}=A_{\rm node}=1/(\pi\Gamma)$. Thus, decreasing the value of $\Gamma$ does not change the ratio $A_{\rm FS}/A_{\rm deep}=1/2$. 

\begin{figure}[!th]
\includegraphics[width = 8.7 cm]{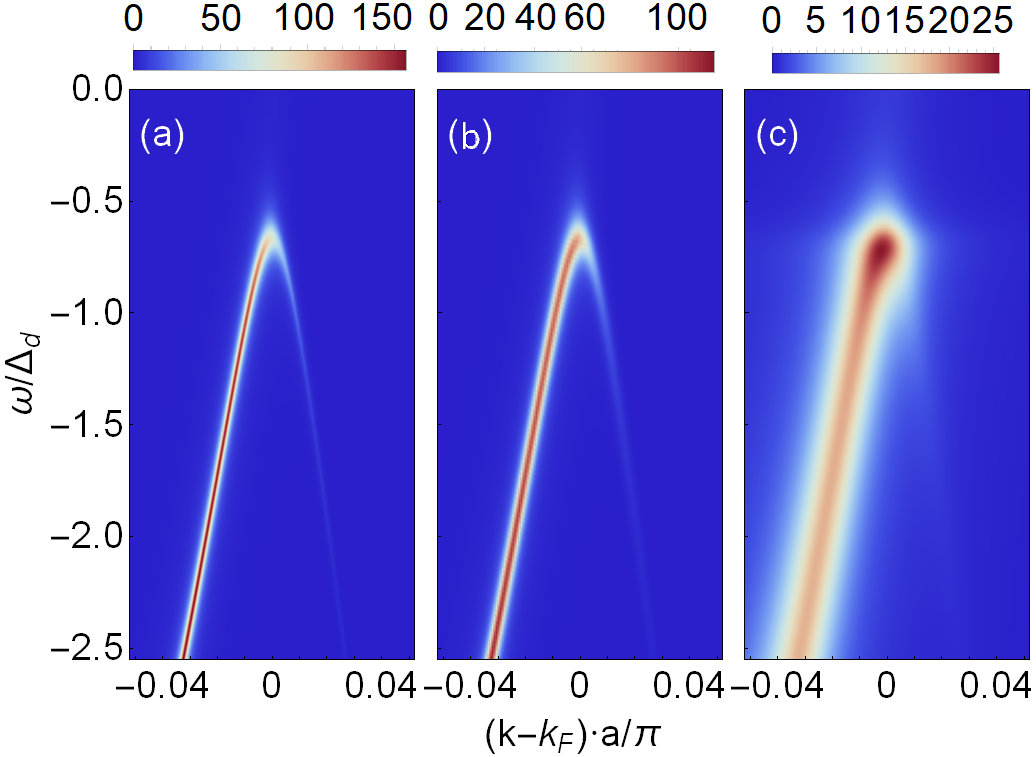}
\caption{Tomographic maps of $A(k,\omega)$ at angle $\theta=22.5^\circ$ for $\Delta_d=30$~meV. (a,b) $\Gamma_s=0$ and $\Gamma=2$~meV. (c) $\Gamma_s=15$~meV  and $\Gamma=3$~meV, as in Fig.~\ref{fig:tomographic_cut}c from the main text. In (b,c) we take into account the finite values of the momentum resolution $\delta k\approx 0.004$~\AA$^{-1}$ and energy resolution $\delta E\approx 1$~meV reported in \cite{Yan23}.}
\label{fig:tomographic_map_suppl}
\end{figure}

Taking into account the finite momentum and energy resolution of the ARPES experiment, for $\Gamma_s=0$ one can estimate $\widetilde{A}_{\rm FS}=\widetilde{A}_{\rm tip}$ and $\widetilde{A}_{\rm deep}=\widetilde{A}_{\rm node}$, where  $\widetilde{A}_{\rm tip}$
and $\widetilde{A}_{\rm node}$ are given by Eq.~\eqref{eq:convolutions}. As a result, the ratio $\widetilde{A}_{\rm FS}/\widetilde{A}_{\rm deep}$ does increase, but Figure~\ref{fig:tomographic_map_suppl}b shows that for $\Gamma_s=0$ the bright spots are still not present. On the other hand, in Fig.~\ref{fig:tomographic_map_suppl}c we show that, for realistic MRDP parameters with finite $\Gamma_s$, bright spots are present also if finite momentum and energy resolutions are taken into account.

\begin{thebibliography}{99}
\bibitem{note}More precisely speaking, the inelastic part of $\Gamma_s(T)$ in a singlet superconductor is a measure of the sum of forward and $2k_F$ scattering.
\end{thebibliography}
\end{document}